\documentclass[manuscript]{acmart}
\AtBeginDocument{%
  }

\usepackage{amsmath}          % For advanced math formatting
\usepackage{graphicx}         % For including graphics
\usepackage{multirow}         % For multirow in tables
\usepackage{booktabs}         % For better looking tables
\usepackage{tabularx}         % For tables with flexible column widths
\usepackage{tabulary}         % For tabular environments
\usepackage{ragged2e}         % For justified text alignment
\usepackage{subcaption}       % For subfigures/subtables
\usepackage{threeparttable}   % For tables with notes
\usepackage{subfiles}         % For multi-file documents
\usepackage{makecell}         % For more table formatting options
\usepackage{siunitx}          % For consistent SI units formatting

\usepackage{caption}
\usepackage{listings}
\usepackage{placeins}

\begin{document}

\title{ChatGPT vs Social Surveys: Probing Objective and Subjective Silicon Population}

\author{Muzhi Zhou}
\affiliation{%
  \institution{The Hong Kong University of Science and Technology (Guangzhou)}
  \city{Guangzhou}
  \country{China}
}
\email{mzzhou@hkust-gz.edu.cn} 

\author {Lu Yu}
\affiliation{%
  \institution{The Hong Kong University of Science and Technology (Guangzhou)}
  \city{Guangzhou}
  \country{China}
}
\email{lyu349@connect.hkust-gz.edu.cn}

\author {Xiaomin Geng} 
\affiliation{%
  \institution{The Hong Kong University of Science and Technology (Guangzhou)}
  \city{Guangzhou}
  \country{China}
}
\email{xgeng312@connect.hkust-gz.edu.cn} 

\author {Lan Luo}
\affiliation{%
  \institution{The Hong Kong University of Science and Technology (Guangzhou)}
  \city{Guangzhou}
  \country{China}
}
\email{lluo476@connect.hkust-gz.edu.cn}

\begin{abstract}
  Recent discussions about Large Language Models (LLMs) indicate that they have the potential to simulate human responses in social surveys and generate reliable predictions, such as those found in political polls. However, the existing findings are highly inconsistent, leaving us uncertain about the population characteristics of data generated by LLMs. In this paper, we employ repeated random sampling to create sampling distributions that identify the population parameters of silicon samples generated by GPT. Our findings show that GPT's demographic distribution aligns with the 2020 U.S. population in terms of gender and average age. However, GPT significantly overestimates the representation of the Black population and individuals with higher levels of education, even when it possesses accurate knowledge. Furthermore, GPT's point estimates for attitudinal scores are highly inconsistent and show no clear inclination toward any particular ideology. The sample response distributions exhibit a normal pattern that diverges significantly from those of human respondents. Consistent with previous studies, we find that GPT's answers are more deterministic than those of humans. We conclude by discussing the concerning implications of this biased and deterministic silicon population for making inferences about real-world populations.
\end{abstract}


\maketitle

\section{INTRODUCTION}

Large Language Models' (LLMs) impressive understanding of language and ability to produce contextually relevant responses have opened up new possibilities for using LLMs to simulate human behavior and to develop or validate research hypotheses about human interactions and perceptions~\cite{science.adi1778, horton2023large, manning2024automated, bail2024can, XU2024103665}. Recently, more studies have taken a progressive approach by utilizing the role-play characteristics of LLMs to substitute human participants in empirical research. In this approach, researchers have examined whether LLMs, being assigned a specific character with pre-defined sociodemographic characteristics or personality traits, exhibit behaviors similar to that of humans~\cite{argyle2023out, aher2023using, Bisbee_Clinton_Dorff_Kenkel_Larson_2024, kim2024aiaugmentedsurveysleveraginglarge, chen2024personapersonalizationsurveyroleplaying, sarstedt2024using}. These roles designated by researchers and deployed in the LLM systems are referred to as the ``homo silicus''~\cite{horton2023large} or the ``silicon samples~\cite{argyle2023out}.'' 

%The exceptional language capabilities of Large Language Models (LLMs) have inspired numerous attempts aimed at understanding human behaviors through artificial agents, as well as gaining insights into LLMs themselves by drawing from human societal contexts. LLMs are models extensively trained on vast amounts of data from diverse sources, including public information, licensed third-party data, and content generated by human reviewers and users, such as books, articles, and web pages~\cite{achiam2023gpt}. Unlike humans, who engage in conversations through intention, reasoning, and inference, LLMs generate sequences of words by selecting the most appropriate next word based on trained statistical models. For example, GPT-3.5 is a conversation-based LLM trained using Reinforcement Learning from Human Feedback (RLHF)~\cite{ouyang2022traininglanguagemodelsfollow,christiano2017deep}. As a result of this training, ChatGPT has demonstrated language abilities that closely resemble those of humans~\cite{strachan2024testing,hagendorff2023human}. Beyond working as an enhanced research tool to improve the efficiency and productivity of academic output~\cite{peterson2024ai}, investigating the ability of LLMs to generate hypotheses through the researcher-machine interaction~\cite{ludwig2024machine,zhou2024hypothesis, NBERw32381}. 

 The alignment between LLM responses and human responses to social survey-like questions remains highly inconsistent. LLM agents have shown the capability to predict outcomes of American presidential polling~\cite{argyle2023out, ShapengJiang2024}. However, the accuracy of these predictions is significantly influenced by the attitudinal information provided~\cite{yang2024largelanguagemodelsllms} and substantial weighting adjustments~\cite{ShapengJiang2024}. Extensive training using interview scripts is also necessary for reasonable performance of LLMs~\cite{park2024generativeagentsimulations1000}. These factors highlight a considerable mismatch between responses from these generally trained LLM agents and those of humans.

 In this paper, rather than estimating the percentage of accuracy in GPT's prediction in a social survey setting, we aim to understand the population characteristics in the eyes of LLMs. Without a clear understanding of the basic sociodemographic and attitudinal features of the responses generated by LLMs, it is challenging to evaluate their potential as a \textit{"silicon population"} to assist our understanding of the human society.

We focus on two key population characteristics. The first includes more objective socioeconomic measures such as sex, age, race, education, and income. The second encompasses more subjective, value-related features, including attitudes toward income inequality and gender roles. Specifically, we aim to determine the population patterns derived from aggregated "homo silicus". The research questions are:

\begin{itemize}

\item{\textbf{RQ1: What is the distribution of socio-demographic characteristics in the population built on responses generated by ChatGPT?}}

\item{\textbf{RQ2:What is the distribution of attitudinal characteristics in the population built on the responses generated by ChatGPT?}}

\end{itemize}

Motivated by the significant potential of LLMs to simulate human respondents for collecting social survey-like information, as well as the current mixed results in this area, we examine and compare both the objective and subjective aspects of data generated by ChatGPT. We use population census and probability social survey data as benchmarks for our analysis. To assess ChatGPT's perception of the U.S. population, we instruct it to draw multiple random samples. The estimates from these samples create a sampling distribution, with a bell shape and the mean representing the population parameter of the GPT-generated data. We then compare this silicon-based population parameter to values derived from the population census and social surveys. This comparison allows us to evaluate how closely the estimates generated from the GPT population align with those from official census data and social surveys. This comparison can uncover potential biases within the LLM's silicon population. The findings have significant implications for research that seeks to utilize aggregated data generated by LLM agents as substitutes for human social surveys in understanding human behaviors and attitudes. Specifically, these results highlight the limitations and potential biases inherent in using LLM-generated data to draw conclusions about real-world populations.

\section{RELATED WORK}

\subsection{Probability Surveys and Inference for Population Characteristics}
Data from probability surveys serve as the foundation for making inferences about a population, particularly when it is impractical or costly to collect information from the entire population of interest~\cite{sarantakos2017social}. Probability sampling ensures that each individual in the target population has a non-zero and equal chance of being selected for the survey sample. This approach allows a small group of individuals in the sample to effectively represent a much larger population. One of the most simple probability sampling methods is simple random sampling.  In this sampling method, each member of the population has an exactly equal chance of being selected. The \textit{representativeness} of the sample is crucial, as it enables statistical generalization to the target population~\cite{bradley2021unrepresentative}. In line with this principle, if we instruct LLMs to randomly select individuals from a given population, the resulting samples should ideally form a representative sample. This would allow us to draw valid inferences about the characteristics of the LLM-generated population based on the sampled data.

% Conducting high-quality, large-scale social surveys has become increasingly challenging. One major issue is the declining response rates observed in numerous social surveys worldwide. For example, in the United States, the response rates for the National Health Interview Survey (NHIS) fell from approximately 95\% in the 1990s to below 75\% by 2014~\cite{fowler2009design}. Low response rates are concerning because they can lead to nonprobability samples, meaning that the characteristics of respondents may differ significantly from those who do not participate. As the nonresponse rate increases, so does the potential for response bias, especially when we cannot identify all factors correlated with both nonresponse and the statistical estimates we aim to achieve~\cite{davern2013nonresponse}.

% Another challenge stems from the increasing length and complexity of questionnaire designs, which can lead to greater measurement errors. Multiple stakeholders, including designers, interviewers, and respondents, invest significant time in the questionnaire design, survey participation, and answering questions~\cite{fowler2009design}. A critical concern arises when respondents encounter complex questionnaires, as this can result in \textbf{measurement errors}. Questions regarding attitudes or beliefs are often lengthy and consist of multiple items, which can be mentally taxing for respondents~\cite{lenzner2010cognitive}. This high cognitive demand can lead to biased responses, answers

\subsection{Role-Playing Language Agents (RPLA) in Answering Survey Questions}

LLMs can exhibit language abilities that align with specific characters and engage in interactions that enhance their performance~\cite{park2023generative, ouyang2022traininglanguagemodelsfollow, aher2023using, wei2022emergent}. This adaptability stems from their ability to process given prompts with information about various personas, behave accordingly, and learn from example demonstrations. A typical Role-Playing Language Agents (RPLA) setup involves prompting LLMs to simulate specific human characteristics and engage in dialogues. Key characteristics often include occupation, gender, and ethnicity. For example, a simple prompt such as "You are a doctor" can effectively guide the LLM's responses.

\subsubsection{Alignment}
Many studies have focused on LLM's ability to answer opinion polls. Argyle et al. (2023) introduced "silicon sampling," in which the GPT-3 language model acts as a proxy for human respondents to fill in voting information ~\cite{argyle2023out}. In this paper, they provided extensive shots to GPT, including party identification and political interest, to predict whether they answered voting for a Republican or Democratic candidate. The tetrachoric correlations of the voting outcomes between GPT and human respondents were all over 0.9, highlighting the ability of LLMs to capture the voting preferences of Americans. Building on this, Sun introduced ``random silicon sampling,'' which assigns demographic distributions from a population to RPLAs to assess partisan attitudes~\cite{sun2024random}. Sun's study found that the generated responses closely mirrored actual U.S. public opinion polls, even for specific social groups, but this is not the case on other non-political attitude-related questions. Bisbee et al, (2024) found that synthetic ChatGPT opinions about their feelings towards different sociopolitical groups look remarkably similar to human American National Election Study respondents with ten given individual traits ~\cite{Bisbee_Clinton_Dorff_Kenkel_Larson_2024}. 

%Similarly, another study found some level of similarities between LLM agents and human respondents in responses to energy social surveys by creating LLM agents with representative characteristics~\cite{ssrnEnergySocial}.

Notably, A follow-up study found that the near-perfect replication of human polling outcomes in the work by Argyle et al. (2023) ~\cite{argyle2023out} is largely due to the assigned features of the RPLA in their study, which overlap with voting outcomes, such as self-identification of political ideology and party affiliation. When these two shortcuts were removed, the performance of GPT-3.5 in accurately predicting polling outcomes declined from the over 90\% level in Argyle et al.'s work to slightly over 60\% ~\cite{yang2024largelanguagemodelsllms}.

%In addition, an increasing number of studies are exploring other potential applications of large language models like ChatGPT in the social sciences. For example, Grossmann has noted that in psychological research, LLMs could act as confederates, providing consistent responses to participants\cite{grossmann2023ai}. However, LLMs are not universally applicable in all contexts. They may be most useful as participants when studying specific topics, using specific tasks, at specific stages of research, or when simulating specific samples. For instance, LLMs might be suitable for completing long surveys, as such surveys risk losing people's attention, but LLMs can rapidly answer hundreds of questions without fatigue~\cite{dillion2023can}. This concept includes "algorithmic fidelity," referring to the model's ability to accurately reflect human thoughts, attitudes, and sociocultural contexts. 

%Finally, A study found that ChatGPT demonstrates a significant and systematic political bias towards the Democratic Party of the United States, Brazil's Lula, and the Labour Party of the United Kingdom \cite{motoki2024more}.

%Similarly, A study prompted the model with various scenarios, such as car buying negotiations or election result predictions, to provide individual recommendations. It was found that these recommendation systems systematically disadvantage names typically associated with racial minority groups and women\cite{haim2024s}. Moreover, a study conducted by Bloomberg found that in recruitment, ChatGPT does not treat all resumes equally. For instance, 

\subsubsection{Mis-alignment}
An underlying assumption in the optimistic exploration of using RPLAs as proxies for human participants is "algorithmic fidelity," which refers to the model's ability to reflect human thoughts, attitudes, and sociocultural contexts accurately. However, since LLMs are trained to respond rather than ensure accuracy, they have a significant potential to produce flawed answers. Their outputs are not grounded in social intelligence or a theory of mind~\cite{messeri2024artificial, LOPEZESPEJEL2023100032}. Furthermore, due to limitations in their training data and safety alignment measures, LLMs may exhibit certain characteristics, such as specific personalities or demographic biases, that do not accurately represent the average population~\cite{santurkar2023whose,huang2023humanity,yuan2023gpt}.

Studies noted that great differences exist between the performance of RPLA and humans. First, the objective world of the LLMs deviates from the real population. Santurkar et al. (2023) found that LLMs tend to reflect the views of younger individuals with higher levels of education~\cite{santurkar2023whose}. The subjective values of LLMs also differ from those of the real population. For example, in terms of political attitudes, LLM responses tend to be more liberal compared to the general population~\cite{rutinowski2024self, hartmann2023political}. Additionally, research has shown that there is a lack of alignment between LLM responses and human respondents when predicting opinions on a broader range of non-political issues~\cite{sun2024random}. This discrepancy highlights the urgent need for extensive fine-tuning of LLMs. Researchers have proposed various methods to enhance the alignment of LLM outputs with human responses~\cite{kim2024aiaugmentedsurveysleveraginglarge, bakker2022fine, scherrer2024evaluating,veselovsky2023generatingfaithfulsyntheticdata, chuang2024demographicsaligningroleplayingllmbased}. However, even when LLMs are provided with comprehensive human interview scripts, their performance in answering questions from general social surveys remains inferior to their performance on Big Five Personality assessments and Economic Behavior Games~\cite{park2024generativeagentsimulations1000}. This finding underscores the considerable challenges in achieving high accuracy for social survey responses compared to other types of human simulations. 

 Furthermore, studies have observed that the variation in responses generated by LLMs is significantly smaller than that of human responses. For example, Bisbee et al. regarding feeling thermometer scores for 11 sociopolitical groups noted that the variations in scores generated by ChatGPT are considerably smaller than the actual scores from the population ~\cite{Bisbee_Clinton_Dorff_Kenkel_Larson_2024}. This narrowing distribution of answers has been noted in several other studies~\cite{argyle2023out, Gordon_2022,kirk2023personalisationboundsrisktaxonomy, mohammadi2024creativityleftchatprice}. Corroborated with this lack of overall variation is the lack of variation across different socioeconomic groups~\cite{machinebias2024}. This highly deterministic nature of LLMs limits our ability to explore and understand the rich diversity in human society. As a result, relying on LLM-generated data may obscure the complexities and variations inherent in human attitudes and behaviors.

\section{METHODOLOGY}

%\begin{multicols}{2}

So far, studies concentrate on predicting public opinions, mostly political ideology or voting outcomes. The alignment in the sociodemographic characteristics between the LLM's world and the real human population remains less studied. We conducted two studies to evaluate GPT's understanding of the objective and subjective human population. In Study 1, we examine the sociodemographic features of the silicon population derived from GPT responses. In Study 2, we investigate attitudes toward income distribution and gender roles of the silicon population based on the demographic data from a population-representative social survey.

\subsection{Model Setting}

We used GPT-3.5-turbo for data generation because of its lower cost and faster processing speeds. We also re-ran the data generation process using GPT-4 but found that its performance was either similar to or even worse than that of GPT-3.5 (a substantial amount of missing values generated), and it operated at a significantly slower speed. The similarities of the performance across models in generating survey-like responses have been noted in other studies~\cite{machinebias2024}.

\begin{figure}[htbp]
    \centering
    \includegraphics[width=1\linewidth]{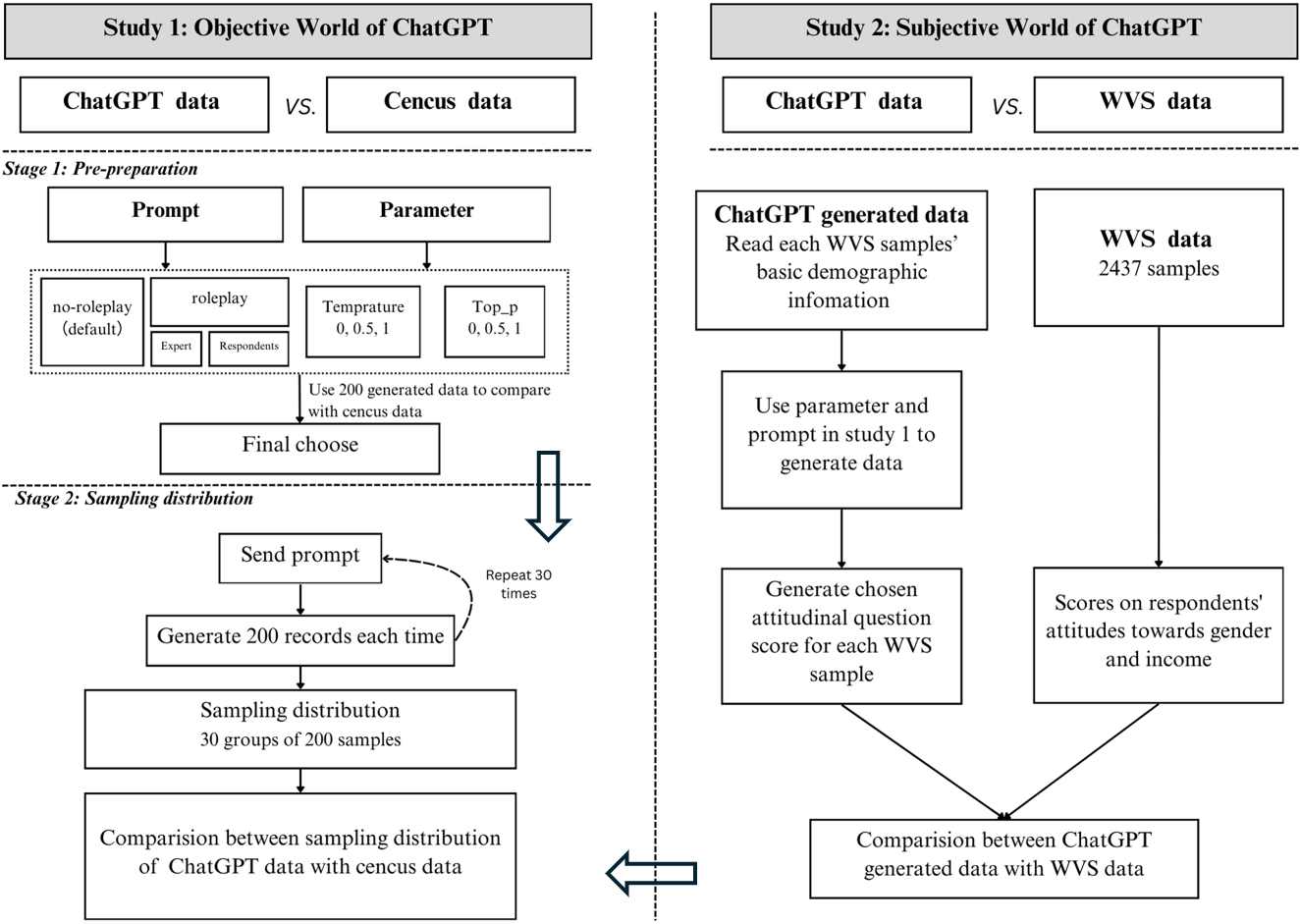}
    \caption{Experiment flow of the two studies}
    \label{flow}
\end{figure}

Model parameters were set to \verb|t|=1, \verb|top_p|=1. We tested various parameter combinations (see Figure \ref{flow}). When \verb|t|=1, ChatGPT provides the most comprehensive demographic feature distribution. For instance, the distribution of race encompasses minority racial groups such as \textit{American Indian and Alaska Native}, and \textit{Two or more categories} when \verb|t|=1 but is missing when \verb|t|=0 or 0.5. We also noted that when  \verb|top_p| equals 0 or 0.5, it exacerbates the issue of missing samples, particularly among minority groups. 

%So, we set the \verb|top_p| value to be 1.

We explored several prompt types, including two role-play settings and one non-role-play setting. In Study 2, we implemented multiple prompt designs and observed that the output improved when responses included text explanations, followed by a score, as typically seen in standard survey questionnaires. 

Figure \ref{flow} presents the experiment flow of the two studies. We conducted multiple tests using various prompts and parameter settings. We selected the prompts and parameters that generate the closest alignment with the benchmark with the least missing values to generate full responses repeatedly. The experiments were conducted multiple times from January to November 2024, and the results are stable.

\subsection{Benchmark Data and Variable Selection} 
For Study 1, we used the 2020 US population census data provided by the United States Census Bureau\href{https://data.census.gov/table?y=2020}{ [link]} as a benchmark. We focus on gender, age, and race. For education, income, and region, where the relevant Census data was unavailable, we relied on the 5-year estimates from the American Community Survey (ACS)\href{https://www.census.gov/programs-surveys/acs}{ [link]}.

For Study 2, we select the \href{https://www.worldvaluessurvey.org/wvs.jsp}{World Values Survey} as our real-world survey data source for answers to subjective questions. The WVS is an international research program conducted every five years, covering a wide range of topics in sociology, political science, economics, social psychology, and so on. It has been active in over 120 societies since 1981. As the largest non-commercial cross-national empirical survey of human beliefs and values, the WVS aims to assist scientists and policymakers in understanding changes in beliefs, values, and motivations globally.

We focus on subjective questions about income inequality and gender roles, the two most common social inequality (income inequality and gender inequality) related measures. In the 2017 wave, the WVS-US dataset contains 2,596 observations. Additionally, we select six basic demographic factors as predictors that may influence attitudes toward income and gender inequality. These factors include gender (Q260), age (Q262), race (Q290), educational attainment (Q275), household income (Q288), party voting (Q223), and region (H\_URBRURAL). 
%Research has shown that these variables are highly effective in shaping individuals' general attitudes toward income distribution and gender inequality.

% Given the need for a large volume of data and the rate limits imposed by ChatGPT, we generated 200 data entries in each batch, repeating this process multiple times until we reached the desired quantity.
%For data generation, we utilized the ChatGPT API to access the ChatGPT model and create data entries in a document. 

% After generating the data, we employed various statistical methods to compare the ChatGPT-generated data with real-world datasets. For comparing means, we primarily used t-tests and ANOVA tests. To compare percentage distributions, we employed the chi-square test. When assessing the data generated by ChatGPT against actual proportions, we relied on a goodness-of-fit chi-square test since we only had information on the distribution of real data without detailed entries. This test helps determine whether the distribution of the generated data aligns with the expected distribution of the actual data. The goodness-of-fit chi-square test is a statistical hypothesis test used to assess whether a variable originates from a specified distribution and is commonly employed to determine if sample data accurately represents the population.

%Specifically, we applied t-tests to assess whether the differences in means between two groups were significant, while ANOVA was used to evaluate the significance of differences in means among multiple groups. 

%\end{multicols}

\subsection{Sampling Distribution} 

 LLMs are known to produce unstable responses, meaning that each sample generated by GPT can vary from one instance to another. This variability is similar to the randomness encountered when selecting a sample from a real population. To address this, we propose constructing a sampling distribution to capture the population parameters within the LLM's framework.

According to the \textit{Central Limit Theorem} (CLT), when we have a set of independent and identically distributed random variables, the distribution of the sample mean will approximate a normal distribution as the sample size becomes sufficiently large, regardless of the original distribution of the variables. This normal distribution will have the same mean as the original distribution, and its variance will be equal to the original variance divided by the sample size (e.g., a sample of 200 data points). By applying this principle, we can calculate the populatoin parameter of the silicon population.

\section{STUDY 1: SILICON POPULATION SOCIODEMOGRAPHICS}\label{study1}       

%\begin{multicols}{2}

We generate a sampling distribution with its mean being the silicon population parameter and compare this mean with the value from a population Census. We consider six demographic variables including gender, age, race/ethnicity, educational attainment, household annual income, and region (urban/rural). 

\subsection{Experiment Settings: Prompts and Model Parameters}

\subsubsection{Prompt}
%We followed the guidelines outlined in the official prompt engineering manual released by OpenAI\footnote{https://platform.openai.com/docs/guides/prompt-engineering}, which emphasizes six strategies, such as \textit{including details in queries to obtain more relevant answers} and \textit{using delimiters to indicate distinct parts of the input among others}. These recommendations were incorporated in crafting our prompts, and 

We used two versions of the prompt: one where ChatGPT generated responses without assuming any role and another called the "role-play prompt." In this version of the prompt, ChatGPT was instructed to assume the role of a survey respondent. We also assigned a role as a survey expert who is good at drawing random samples. The prompt is provided in Github\footnote{https://anonymous.4open.science/r/Surrogate/README.md}. Table \Ref{s1-sampledist} reports one example of the samples generated by GPT. The results are stable across the three prompt types. For the following analysis, we apply the default role prompt, given that there are more values assigned to the tails of the distribution for race, education, and income groups.

\begin{table}[htp]
\centering
\caption{Distribution of one sample with size 200 from three prompt settings}
\label{s1-sampledist}
\resizebox{0.5\textwidth}{!}{
\begin{tabular}{l *{6}{c} l} % Correct column count
\toprule
\multicolumn{1}{l}{\textbf{}} &\multicolumn{1}{l}{\textbf{}} & \multicolumn{1}{c}{\textbf{Respondent}} & \multicolumn{1}{c}{\textbf{Expert}} & \multicolumn{1}{c}{\textbf{Default}} & \textbf{Census} \\
\midrule
\multirow{2}{*}{\textbf{Gender}} & Female  & 50.00\%  & 50.00\%  & 49.00\% & 50.9\% \\ 
& Male & 50.00\% & 50.00\%  & 51.00\% & 49.1\% \\ 
\midrule
\multirow{3}{*}{\textbf{Age}} & Median & 37 & 40 & 38 & 38.8 \\ 
& Min & 18  & 20 & 18 & / \\ 
& Max  & 65  & 70 & 70 & / \\ 
\midrule
\multirow{5}{*}{\textbf{Race}} 
& White  & 40.50\%  & 38.50\%  & 44.00\% & 57.84\% \\
& Black or African  & 22.50\%  & 22.50\%  & 21.00\% & 12.05\% \\ 
& Hispanic or Latino & 20.00\% & 19.50\%  & 18.00\% & 18.73\% \\ 
& Asian & 15.50\% & 17.00\%  & 14.00\% & 5.92\% \\ 
& Others & 1.50\% & 2.5\% & 3.00\% & 5.46\% \\ 
\midrule
\multirow{7}{*}{\textbf{Education}} & 
Less than 9th Grade& 1.50\% & 1.01\%  & 2.50\% & 5.77\% \\ 
& 9th to 12th Grade & / & /  & / & 5.78\% \\ 
& High School Graduate & 21.00\%  & 18.59\% & 21.00\% & 27.32\% \\ 
& Some College, No Degree  & 32.50\% & 31.16\% & 32.50\% & 23.14\% \\ 
& Associate's Degree & 9.50\% & 10.55\% & 10.50\% & 7.60\% \\ 
& Bachelor's Degree & 25.00\%& 26.13\% & 22.00\% & 19.20\% \\ 
& Graduate or Professional & 10.50\% & 12.56\% & 11.50\% & 11.18\% \\ 
\midrule
\multirow{10}{*}{\textbf{Income}} & 
Less than \$10,000 & 1.50\% & 0.50\% & 2.50\% & 5.80\% \\ 
& \$10,000 to \$14,999 & 1.00\% & 1.00\% & 0.50\% & 4.10\% \\ 
& \$15,000 to \$24,999 & 9.00\% & 5.50\% & 8.50\% & 8.50\% \\ 
& \$25,000 to \$34,999 & 15.50\% & 13.00\% & 12.00\% & 8.60\% \\ 
& \$35,000 to \$49,999 & 21.50\% & 21.00\% & 22.00\% & 10.60\% \\ 
& \$50,000 to \$74,999 & 21.50\% & 23.50\% & 25.50\% & 12.30\% \\ 
& \$75,000 to \$99,999 & 13.50\% & 14.50\% & 13.50\% & 9.60\% \\ 
& \$100,000 to \$149,999 & 7.50\% & 8.50\% & 8.00\% & 7.30\% \\ 
& \$150,000 to \$199,999 & 4.50\% & 4.00\% & 5.50\% & 5.40\% \\ 
& \$200,000 or more & 5.50\% & 6.00\% & 4.50\% & 6.30\% \\
\bottomrule
\end{tabular}}
\end{table}

\subsubsection{Between-sample independence}
We perform several iterations of this data generation process. To test the independence between batches, we utilize the same prompt and model parameters (\verb|t|=1, \verb|top_p|=1) to instruct GPT to generate 200 data points (across 10 batches). Subsequently, we employ ANOVA to examine whether the mean differences of all variables in each batch were statistically correlated. The results indicate that there were no significant between-group differences observed across all variables, including gender (F=0, df=9, $p=1$), age (F=0.185, df=9, $p>0.5$), race (F=0.446, df=9, $p>0.5$), educational attainment (F=0.121, df=9, $p>0.5$), income (F=0.767, df=9, $p>0.5$), and region (F=0.444, df=9, $p>0.5$). Post-hoc tests revealed that the differences in mean values between any two groups for all variables were uniformly non-significant. The independence between these different iterations of data generation enables those multiple iterations or batches of data to form a sampling distribution.

%Statistical tests typically require a sample of independent observations, where the value of one observation is not influenced by the value of other observations. In our case, data is generated in batches and then aggregated as a random sample as in survey research. If the data between each batch is not independent (meaning the values in later data depend on the values in the earlier data), it can substantially affect the representativeness of the multi-round generation approach. However, if the data in each batch is independent, the multi-round generation strategy can be seen as repeatedly sampling from a hypothetical sampling frame of a population. 

\subsubsection{Sampling distribution}
We now simulate the simple random sampling statistical process. We treat every 200 data points as a single random sample and repeat redrawing the random sample 30 times from the silicon version of the ``US population in 2020.'' We calculate the mean from each of these 30 iterations and produce a sampling distribution of sample means for demographic characteristics. This process will form a sampling distribution that approaches a normal distribution. Together with the standard deviation of this sampling distribution, we can examine whether the mean from the Census (red vertical line) falls within the 95\% confidence interval of the sampling distribution of the GPT population.

%Therefore, by calculating the sample means through multiple batches of data, we obtain a sampling distribution that approaches a normal distribution. Together with its standard deviation, this helps us to estimate population parameters more accurately and to compare them with data from the census

\subsection{Results}

 We constructed sampling distribution graphs for gender, age, race, region, income, and education level from the GPT population. For categorical variables like race, income, and education with multiple categories, we converted them into dummy variables and created graphs for each to compare with census data in detail. 
 
The sampling distribution graphs for each variable are shown below. Each graph includes reference lines for the 95\% confidence interval (gray line) ($1.96*s.d.$), the mean of the sampling distribution (black line), and the census value (red line). If the census value falls outside the confidence interval, it indicates that the Census value is either higher or lower than the parameter from the GPT population. First, the sampling distributions of the sample mean generally conform to a normal distribution, indicating that the data generated by ChatGPT adhere to the Central Limit Theorem. This confirms that these samples are randomly drawn from the silicon population from GPT. Next, we will provide a detailed interpretation of each variable.

\begin{figure}[hbt!]
    \hspace*{\fill}
     \begin{subfigure}{0.15\textwidth}
        \includegraphics[width=\linewidth]{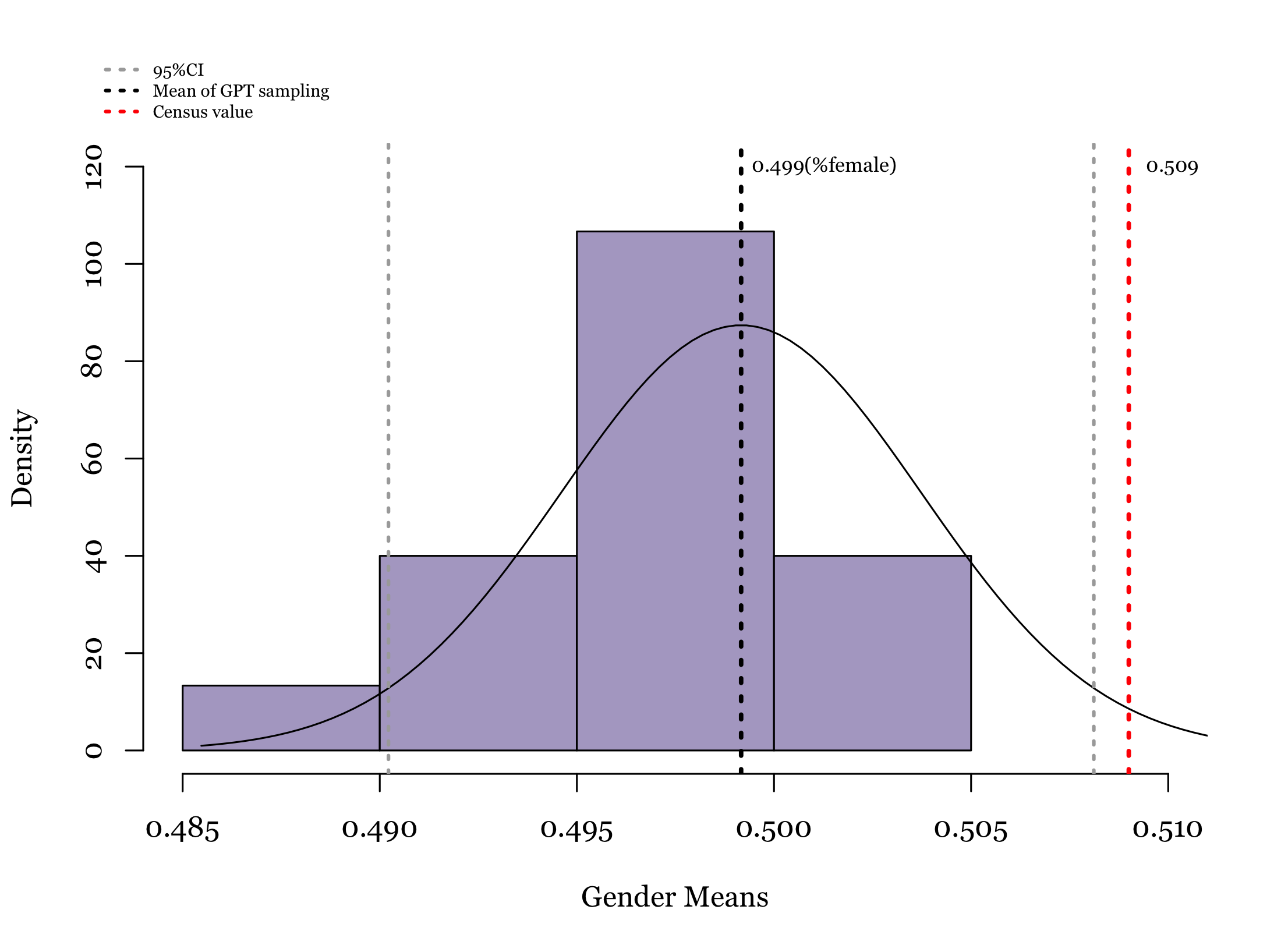}
        \caption{Gender dummy}
    \end{subfigure}  
       \hfill
    \begin{subfigure}{0.15\textwidth}
        \includegraphics[width=\linewidth]{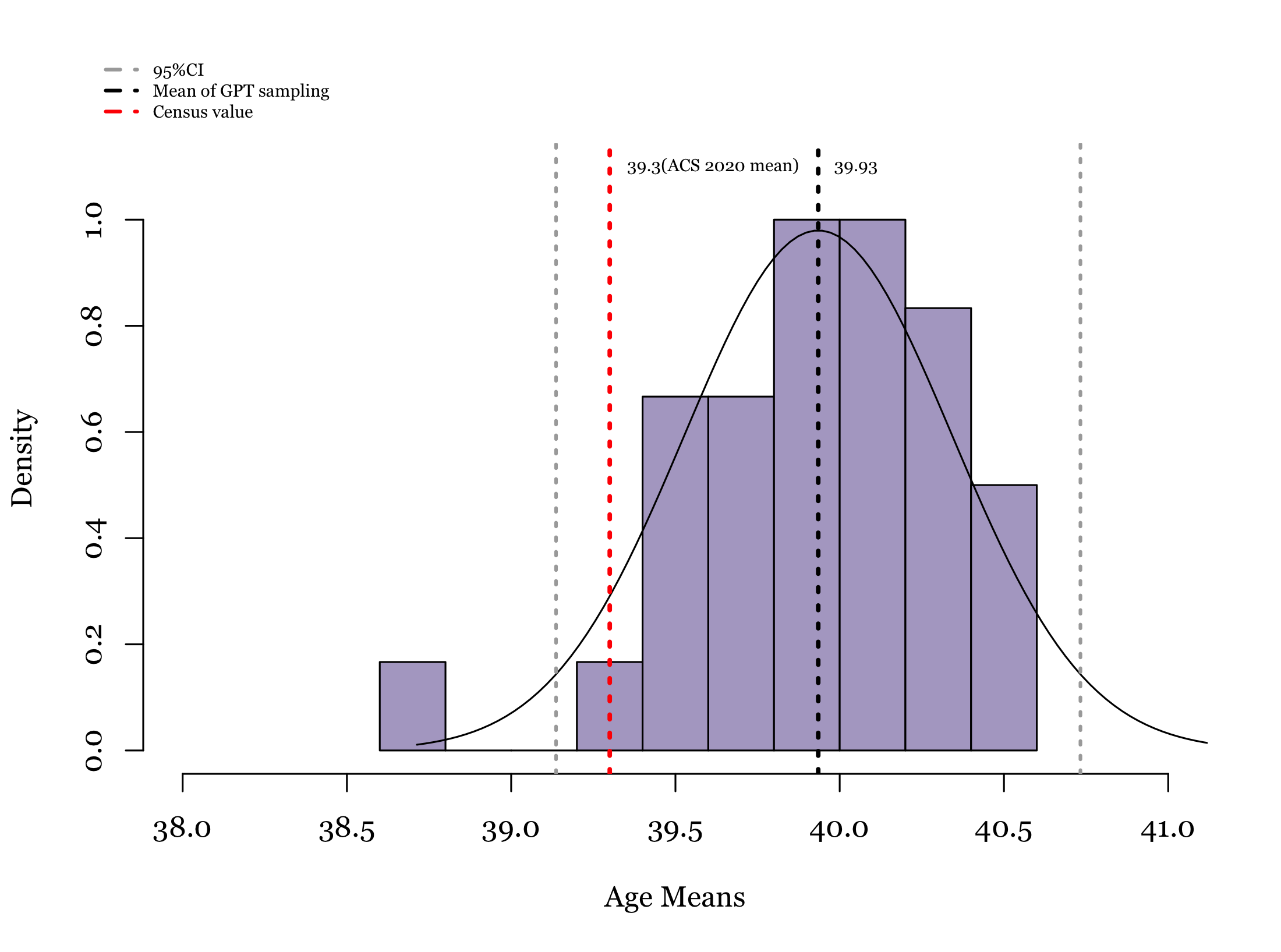}
        \caption{Age}
    \end{subfigure}  
       \hfill
    \begin{subfigure}{0.15\textwidth}
        \includegraphics[width=\linewidth]{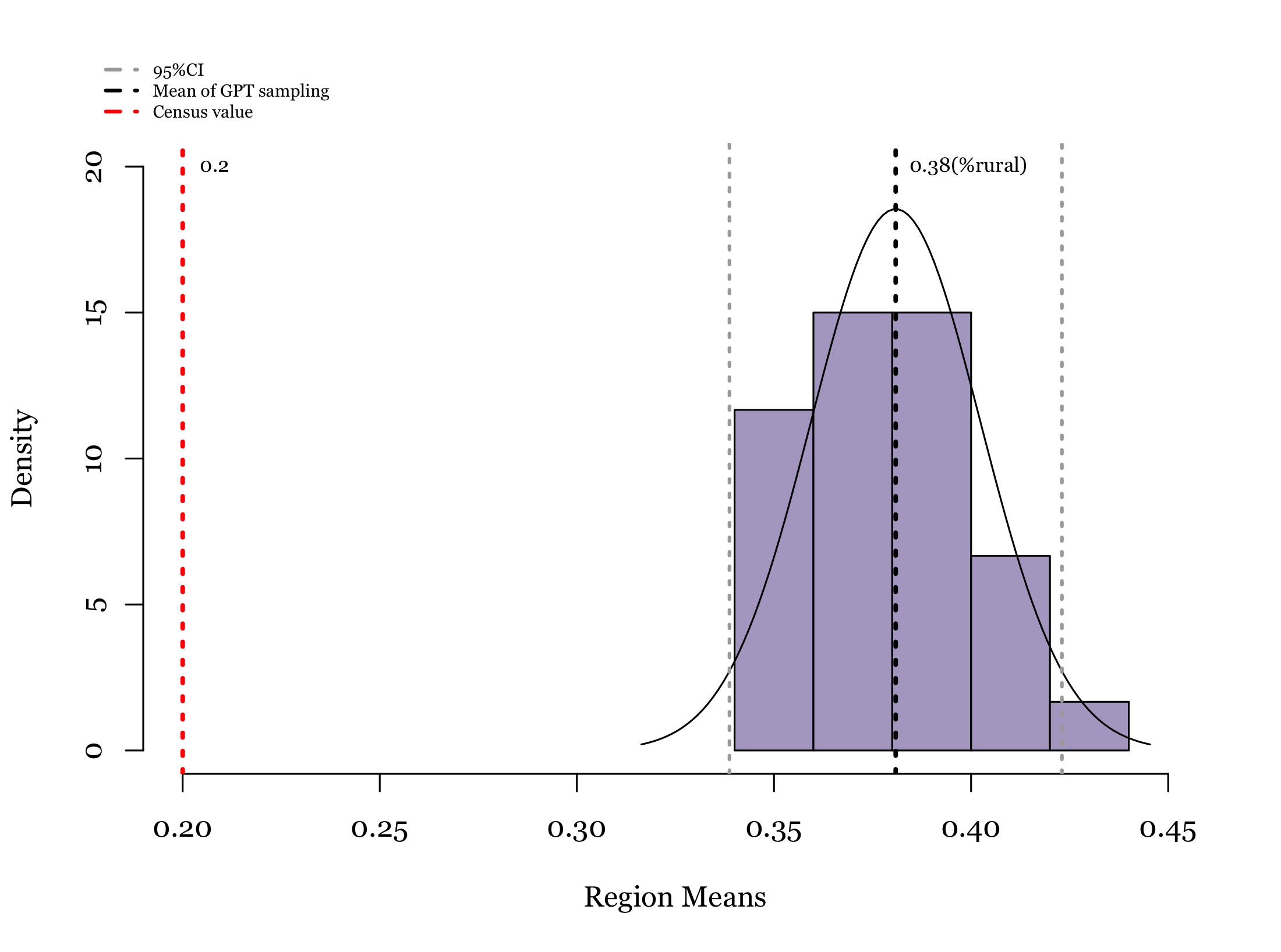}
        \caption{Region dummy}
    \end{subfigure}
    \caption{Sampling distribution of gender, age, and region}
    \label{f1-4}
       \end{figure}

\textbf{1) Gender:} 
For gender, the mean of the sampling distribution is 0.499 (female proportion), while the female proportion from the census data is 0.509 (Figure \ref{f1-4}). The census value does not fall within the confidence interval of the sampling distribution, but the difference is very small.  

    % Subfigures
%     \begin{minipage}{\textwidth}  % Wrap subfigures in a minipage
%        % \centering
%         \begin{subfigure}{0.23\textwidth}
%             \includegraphics[width=\linewidth]{Figures/Study1_Sampling_distribution/age/age1_distribution.png}
%             \caption{age 18-24}
%         \end{subfigure}
%         \hfill
%         \begin{subfigure}{0.23\textwidth}
%             \includegraphics[width=\linewidth]{Figures/Study1_Sampling_distribution/age/age2_distribution.png}
%             \caption{age 25-39}
%         \end{subfigure}
%         \hfill
%         \begin{subfigure}{0.23\textwidth}
%             \includegraphics[width=\linewidth]{Figures/Study1_Sampling_distribution/age/age3_distribution.png}
%             \caption{age 40-59}
%         \end{subfigure}
%         \hfill
%         \begin{subfigure}{0.23\textwidth}
%             \includegraphics[width=\linewidth]{Figures/Study1_Sampling_distribution/age/age4_distribution.png}
%             \caption{age 60 and above}
%         \end{subfigure}
%     \end{minipage}
%     \caption{The sampling distribution of age groups}
%     \label{f1-4}

\textbf{2) Age:} 
The mean age of the sampling distribution is 39.93 years (Figure \ref{f1-4}). We obtained the mean age from the ACS 2020 as a comparable benchmark for comparison. The results indicate that the mean age from ACS 2020 falls within the confidence interval of the sampling distribution. 
%To facilitate further comparison, we categorized age into four groups: 18-24 years, 25-39 years, 40-59 years, and 60 and older years (Figure \ref{f1-4}). The results indicate that ChatGPT's estimate for the proportion of the 18-24 age group falls within the confidence interval, but it overestimates the proportions of the 25-39 and 40-59 age groups and underestimates the proportion of the 60 and above age group. 

\textbf{3) Region:} 
For the region variable, the mean of the sampling distribution is 0.38, compared to 0.2 from the census data (20\% rural residents) (Figure \ref{f1-4}). This suggests a substantial overestimation of the proportion of rural residents in the silicon population.

\begin{figure}[hbt!]
    \hspace*{\fill}
     \begin{subfigure}{0.15\textwidth}
        \includegraphics[width=\linewidth]{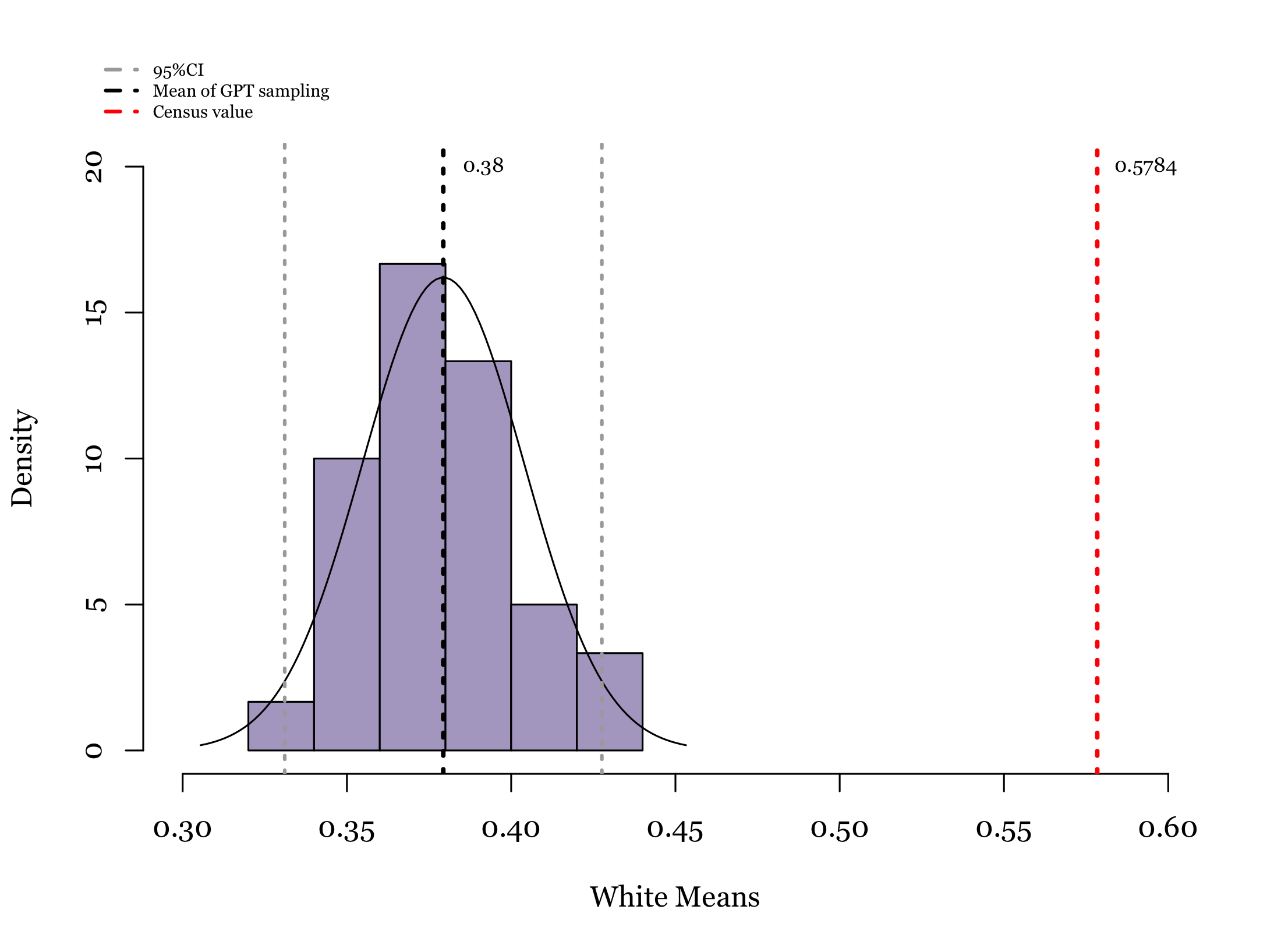}
        \caption{White}
    \end{subfigure}
        \hfill
    \begin{subfigure}{0.15\textwidth}
        \includegraphics[width=\linewidth]{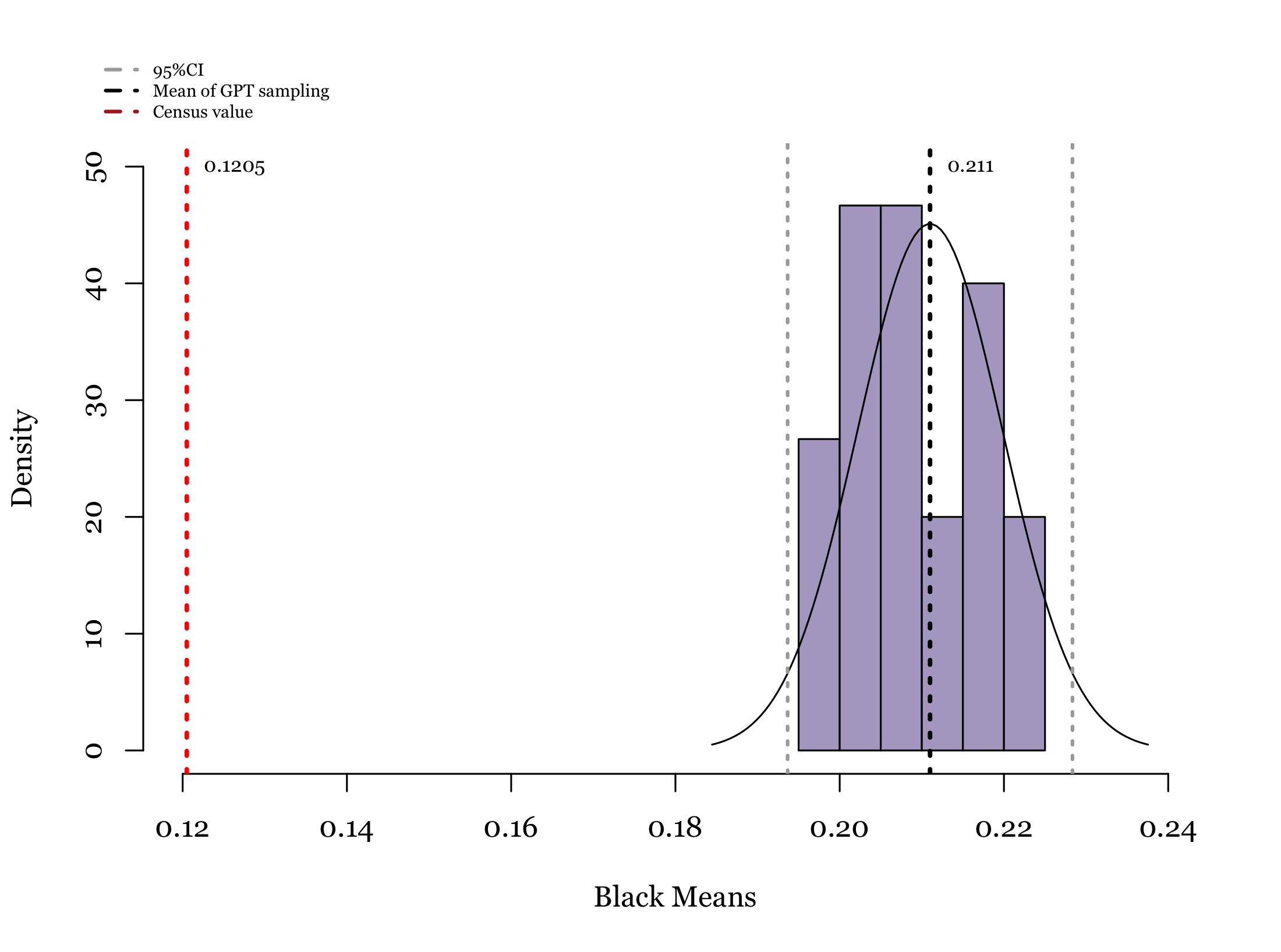}
        \caption{Black}
    \end{subfigure}
     \hfill
    \begin{subfigure}{0.15\textwidth}
        \includegraphics[width=\linewidth]{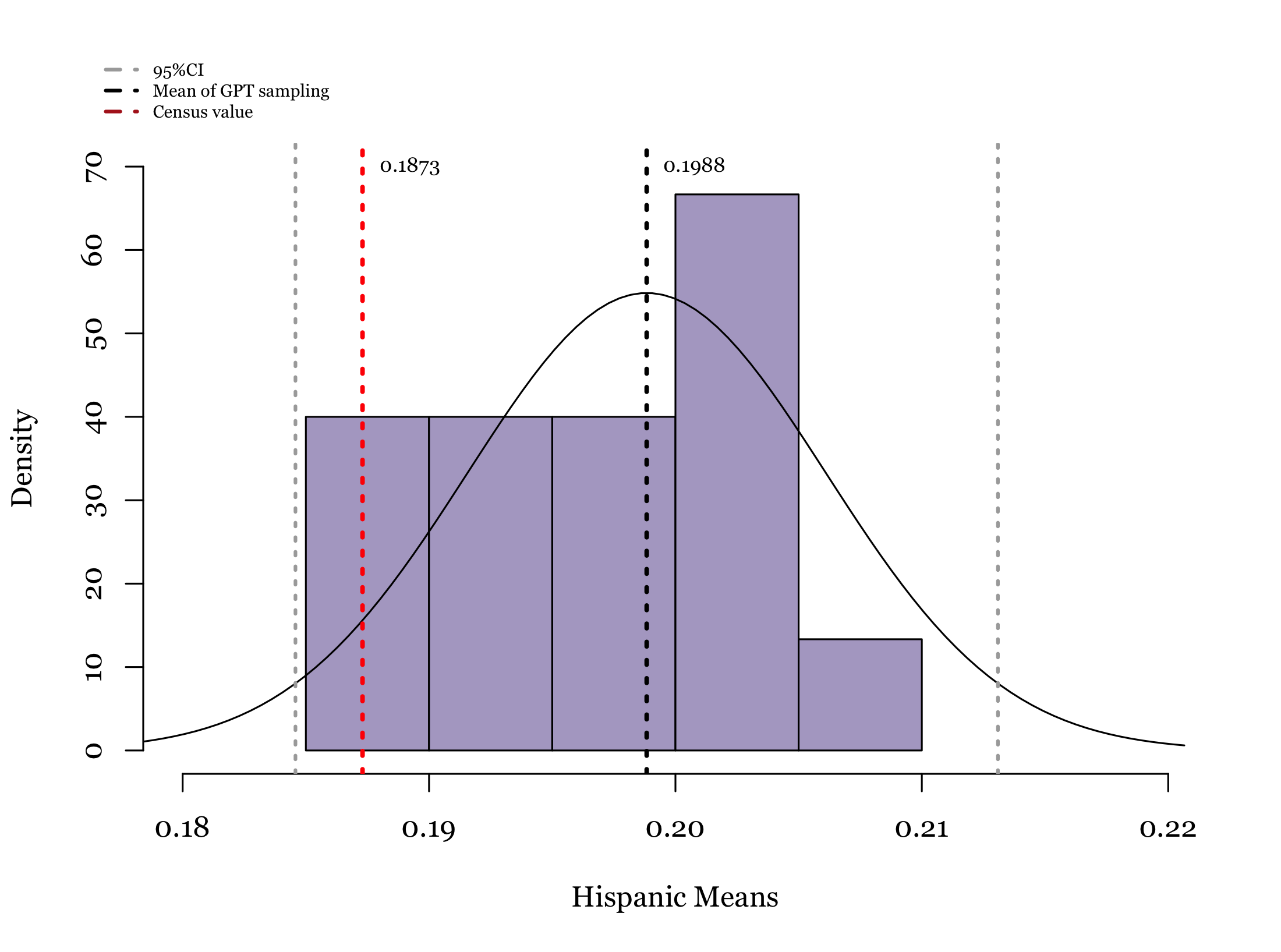}
        \caption{Hispanic}
    \end{subfigure}

     % Row 2
    \vspace{10pt}
    \begin{subfigure}{0.15\textwidth}
        \includegraphics[width=\linewidth]{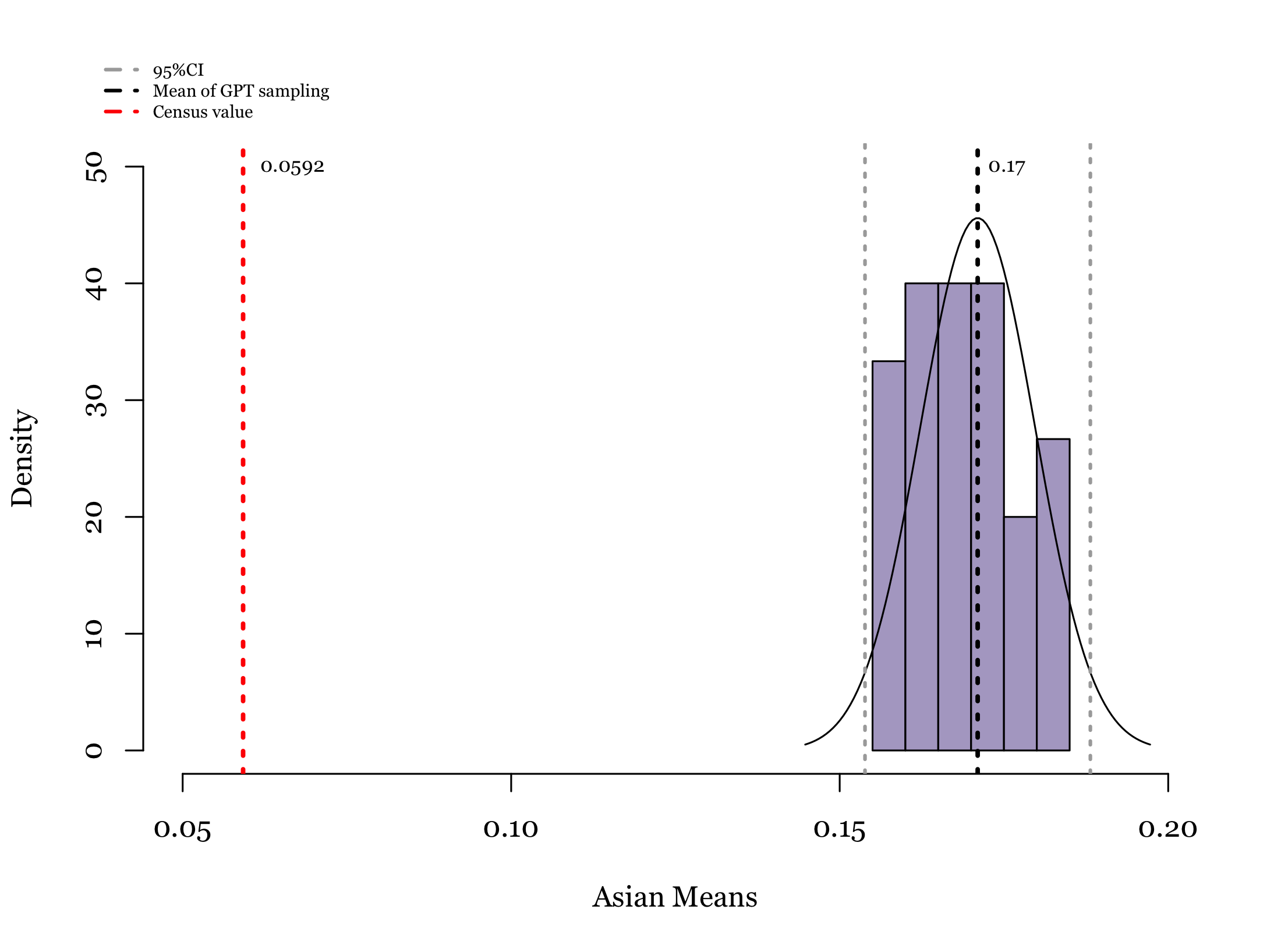}
        \caption{Asian}
    \end{subfigure}
    \hfill
    \begin{subfigure}{0.15\textwidth}
        \includegraphics[width=\linewidth]{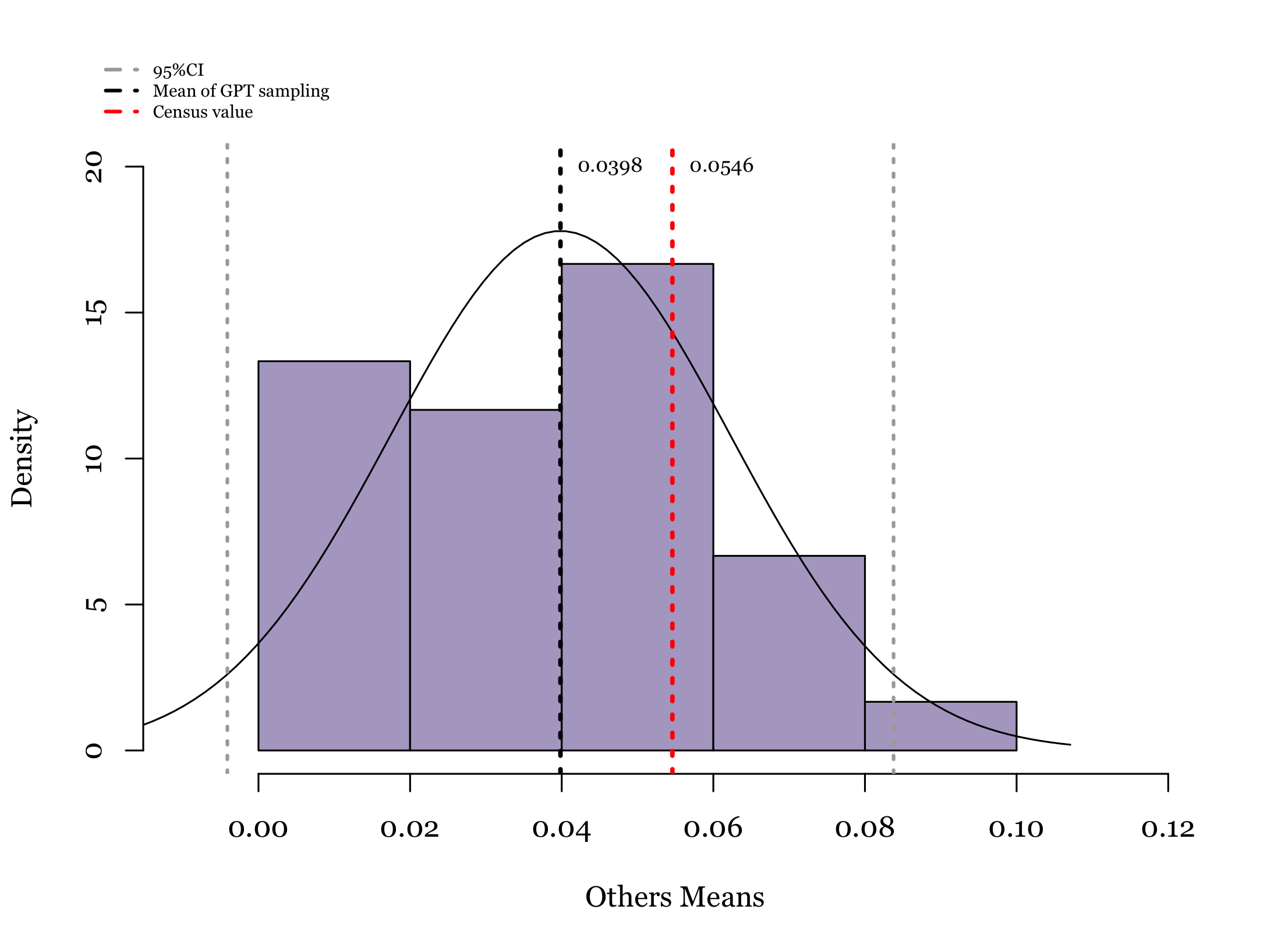}
        \caption{Others}
    \end{subfigure}
    \caption{The sampling distribution of racial groups}
    \label{f1-5}
    \end{figure}

\textbf{4) Race:}
We categorized race into five groups: Asian, Black, Hispanic, White, and Others (Figure \ref{f1-5}). The results indicate that ChatGPT substantially underestimates the proportions of Whites and overestimates the proportions of Blacks and Asians. For Hispanics and Others, the Census values fall within the confidence interval.

\begin{figure}[hbt!]
   \hspace*{\fill}
    \begin{subfigure}{0.15\textwidth}
        \includegraphics[width=\linewidth]{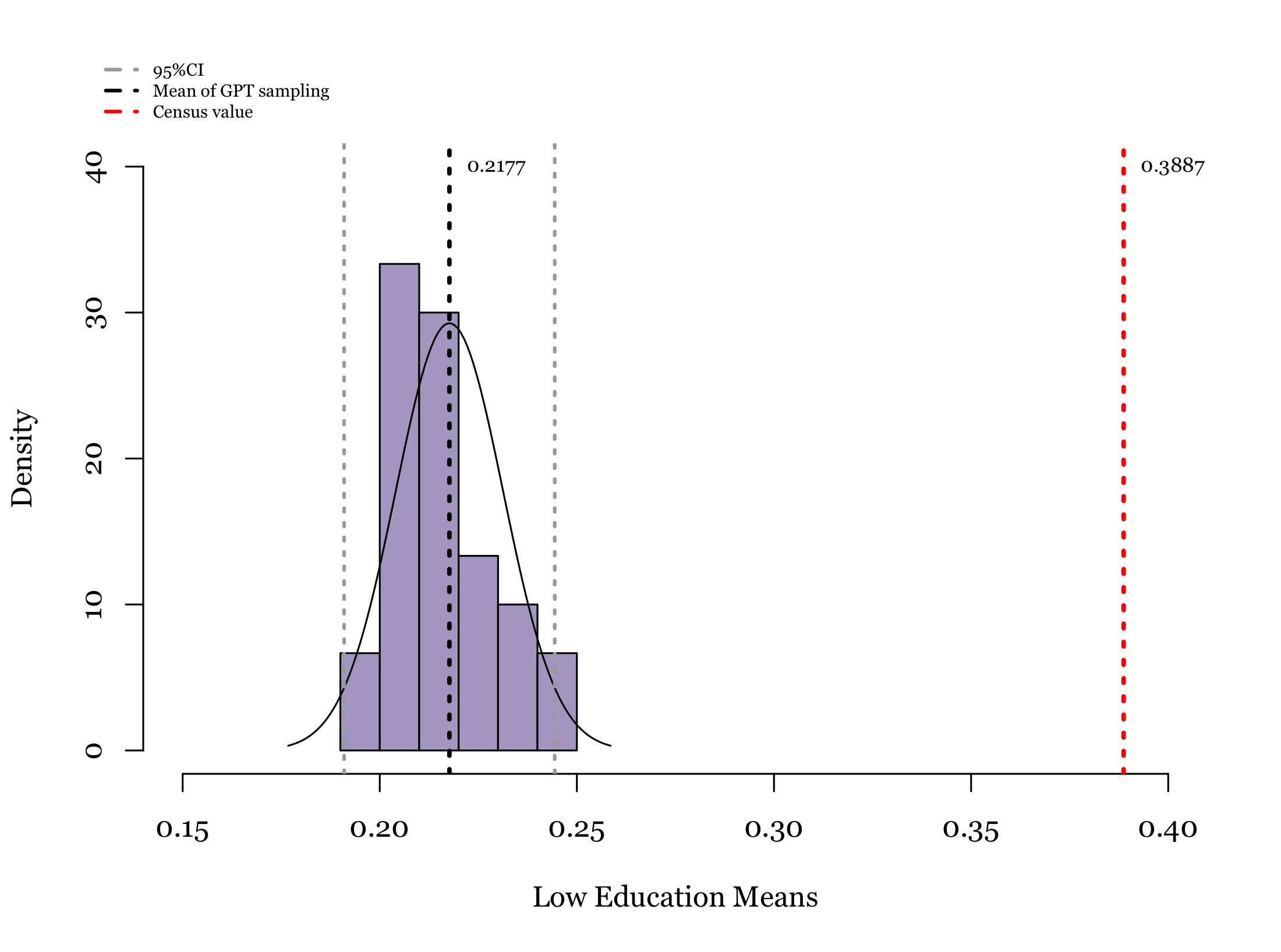}
        \caption{Low education}
    \end{subfigure}
    \hfill
    \begin{subfigure}{0.15\textwidth}
        \includegraphics[width=\linewidth]{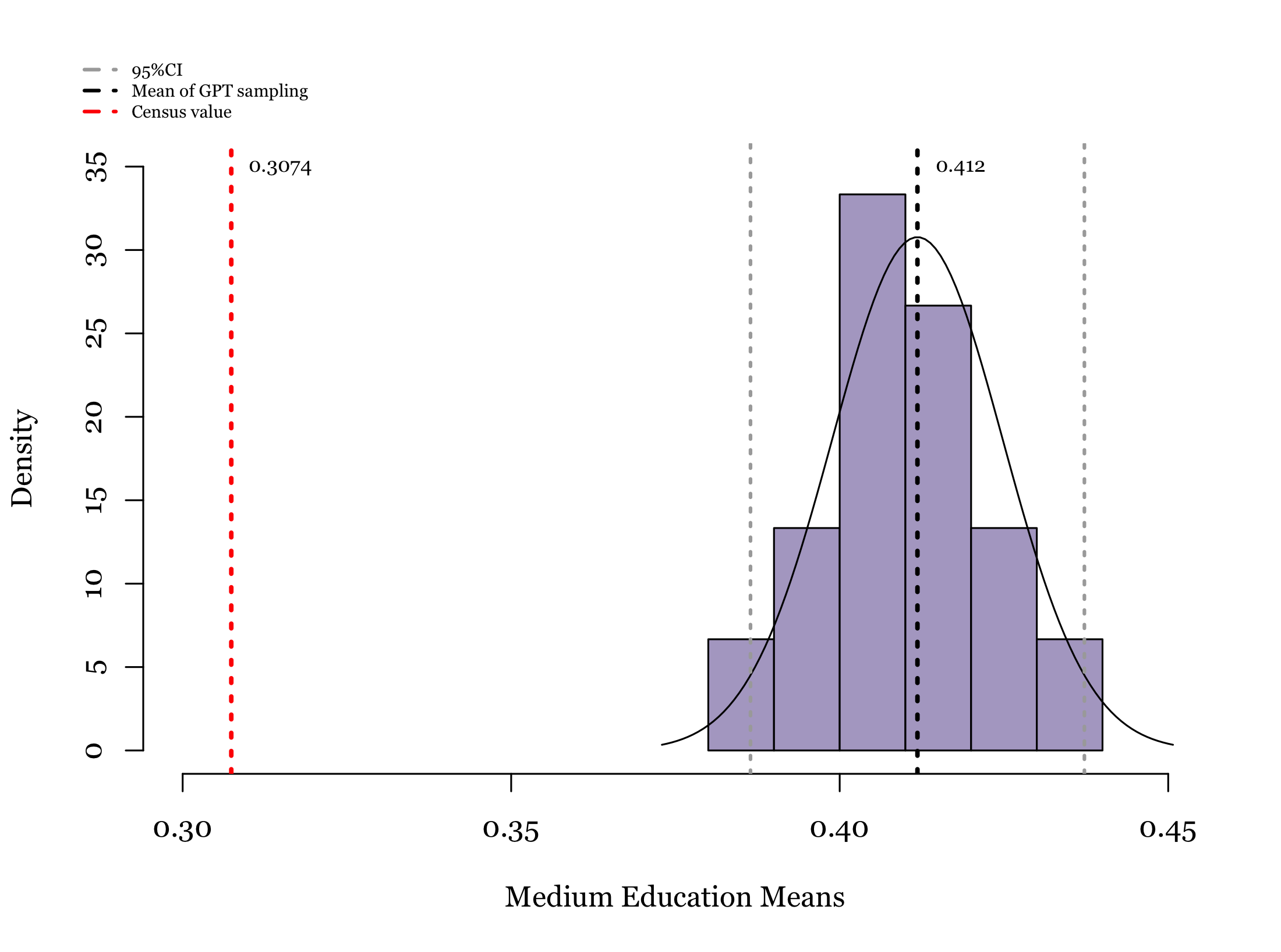}
        \caption{Medium education}
    \end{subfigure}
    \hfill
    \begin{subfigure}{0.15\textwidth}
        \includegraphics[width=\linewidth]{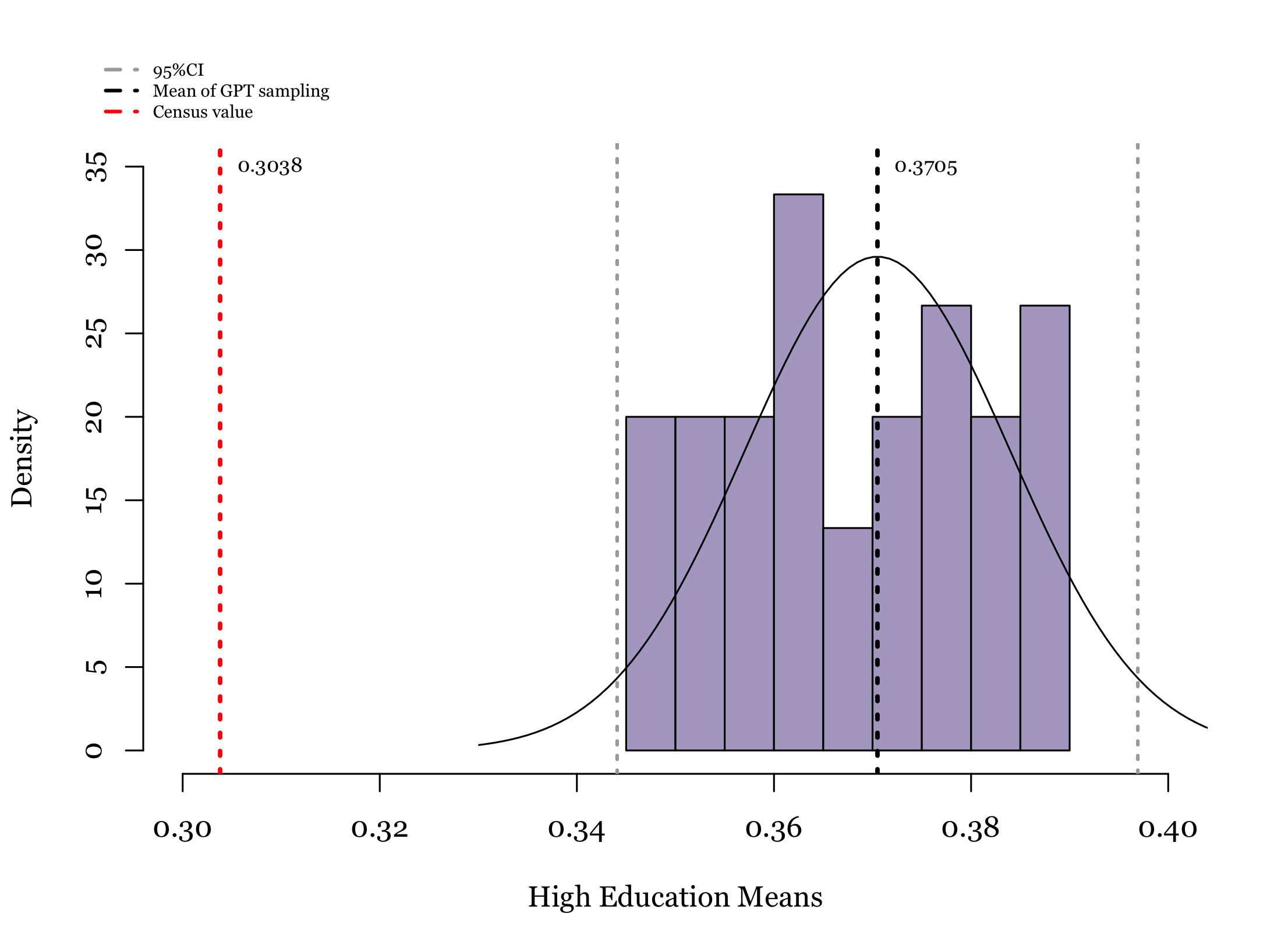}
        \caption{High education}
    \end{subfigure}
    \caption{The sampling distribution of education groups}
    \label{f1-6}
    \end{figure}

\textbf{5) Education}
Education is categorized it into three groups: low, medium, and high (Figure \ref{f1-6}). \textit{Low} refers to high school graduates and below, \textit{medium} includes education levels above high school but below a bachelor's degree (e.g., some college), and \textit{high} represents a bachelor's degree and above. The results indicate that GPT underestimates the proportion of low-education group and overestimates the proportions of medium and high-education groups. 

% \begin{figure*}[hbt!]
%     \centering
%     \includegraphics[width=0.5\textwidth]{Figures/Study1_Sampling_distribution/income/inc_distribution.png} % 将宽度设置为文字宽度的一半
%     \caption{The sampling distribution of income}
%     \label{f1-7}
% \end{figure*}

\begin{figure}[hbt!] 
    % Row 1
    \centering
    \begin{subfigure}{0.15\textwidth}
        \includegraphics[width=\linewidth]{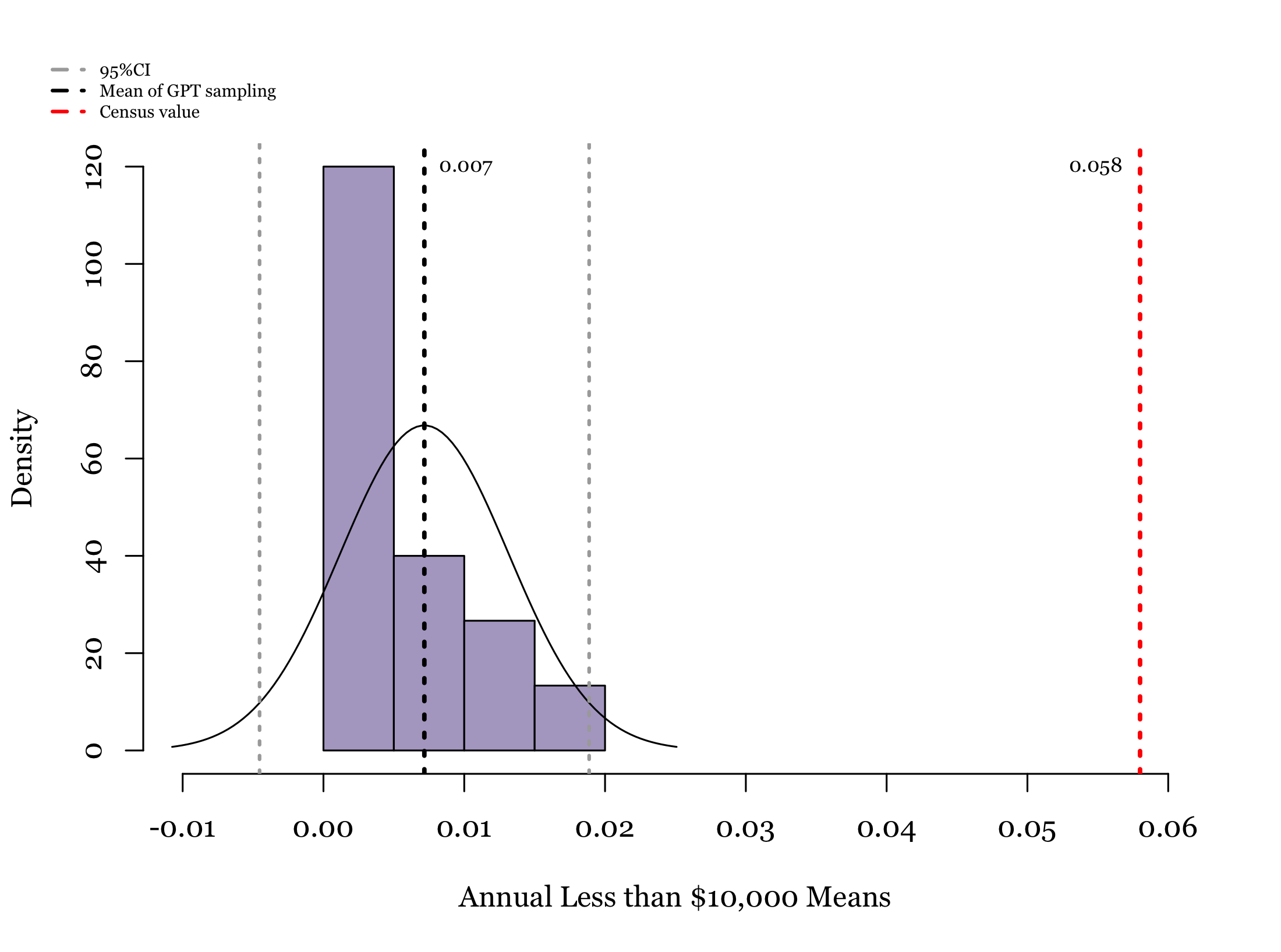}
        \caption{Less than \$10000}
    \end{subfigure}
    \hfill
    \begin{subfigure}{0.15\textwidth}
        \includegraphics[width=\linewidth]{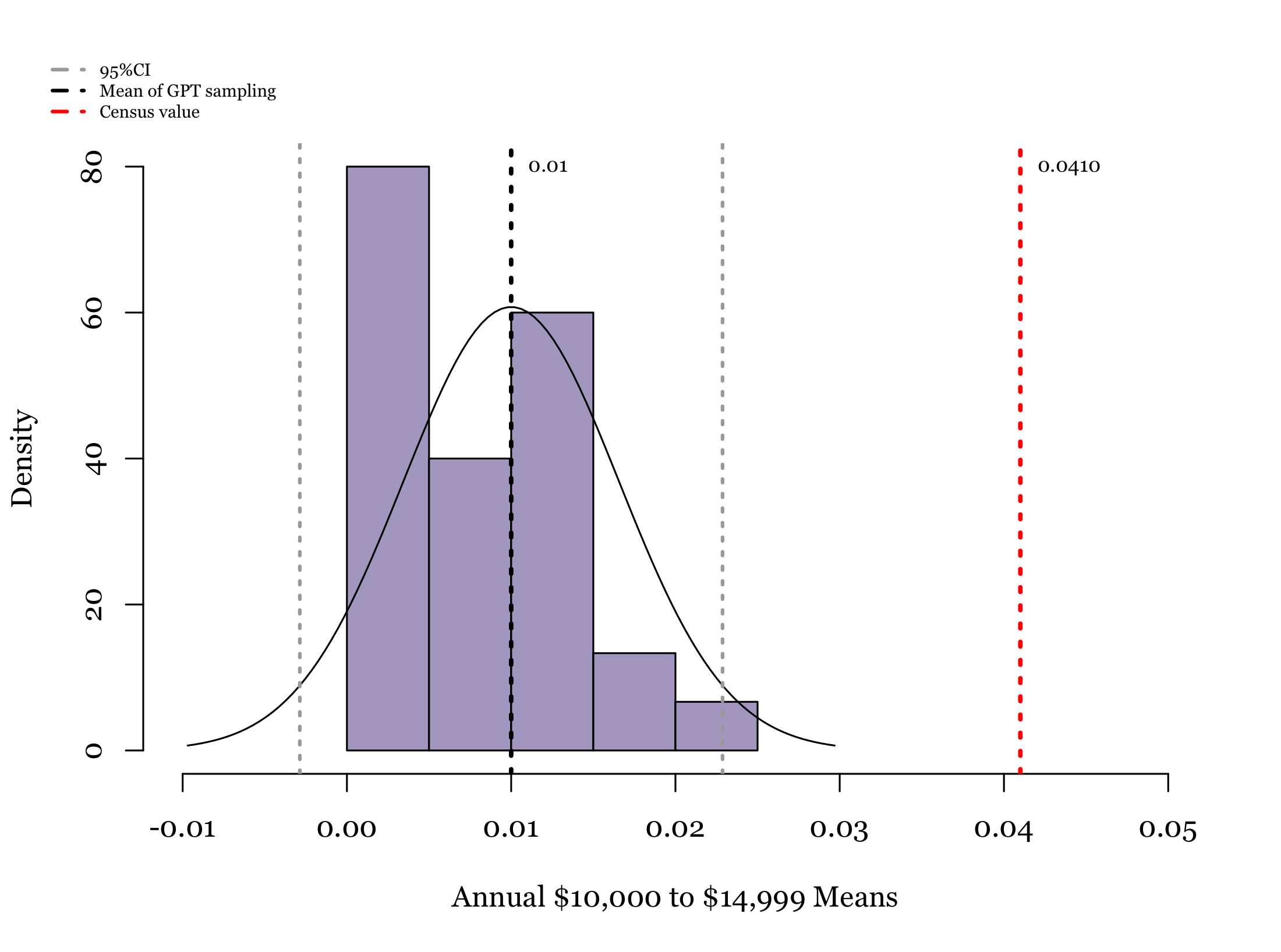}
        \caption{\$10000-\$14999}
    \end{subfigure}
    \hfill
    \begin{subfigure}{0.15\textwidth}
        \includegraphics[width=\linewidth]{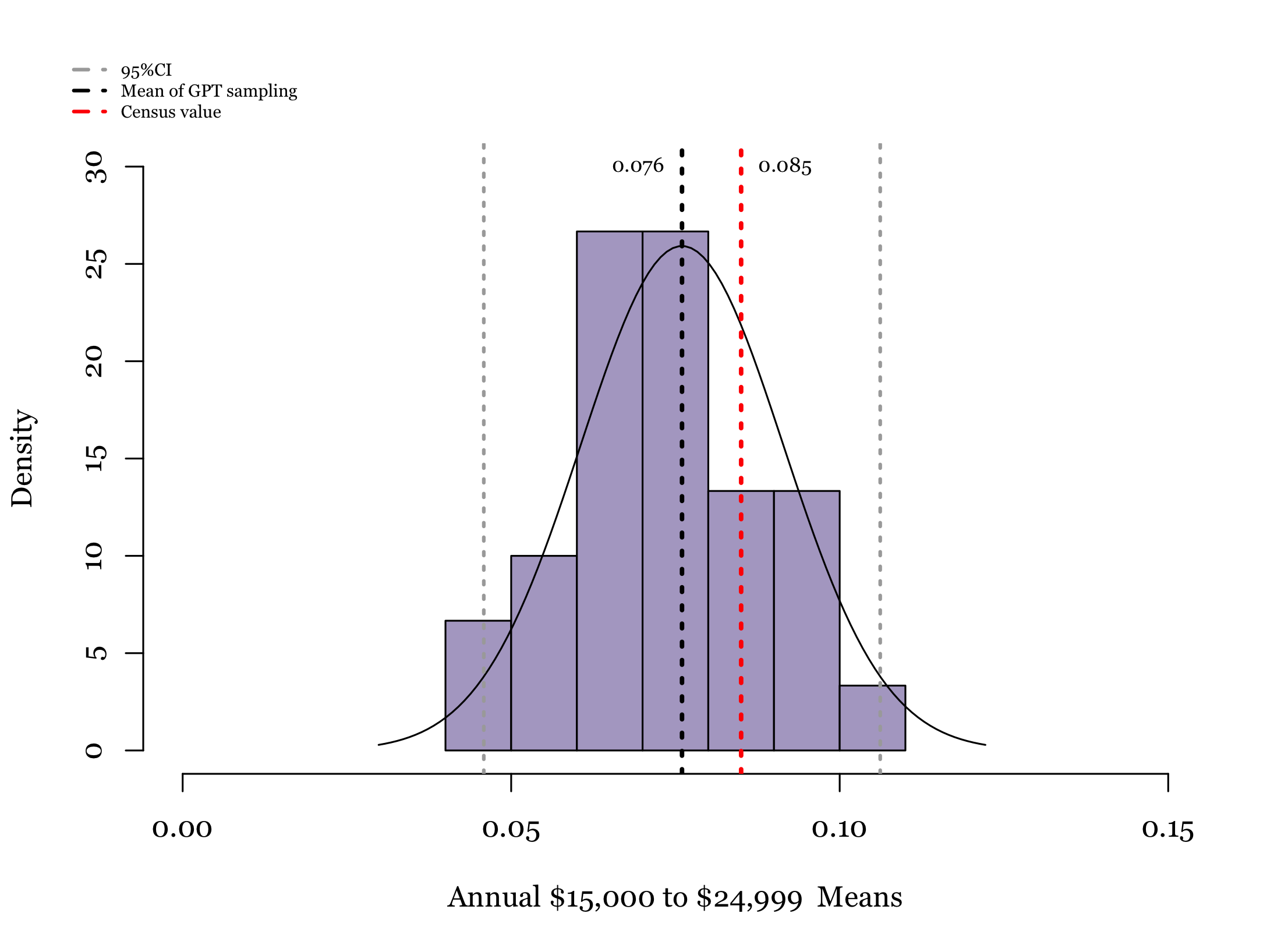}
        \caption{\$15000-\$24999}
    \end{subfigure}

    % Row 2
    \vspace{10pt}  
    \centering
     \begin{subfigure}{0.15\textwidth}
        \includegraphics[width=\linewidth]{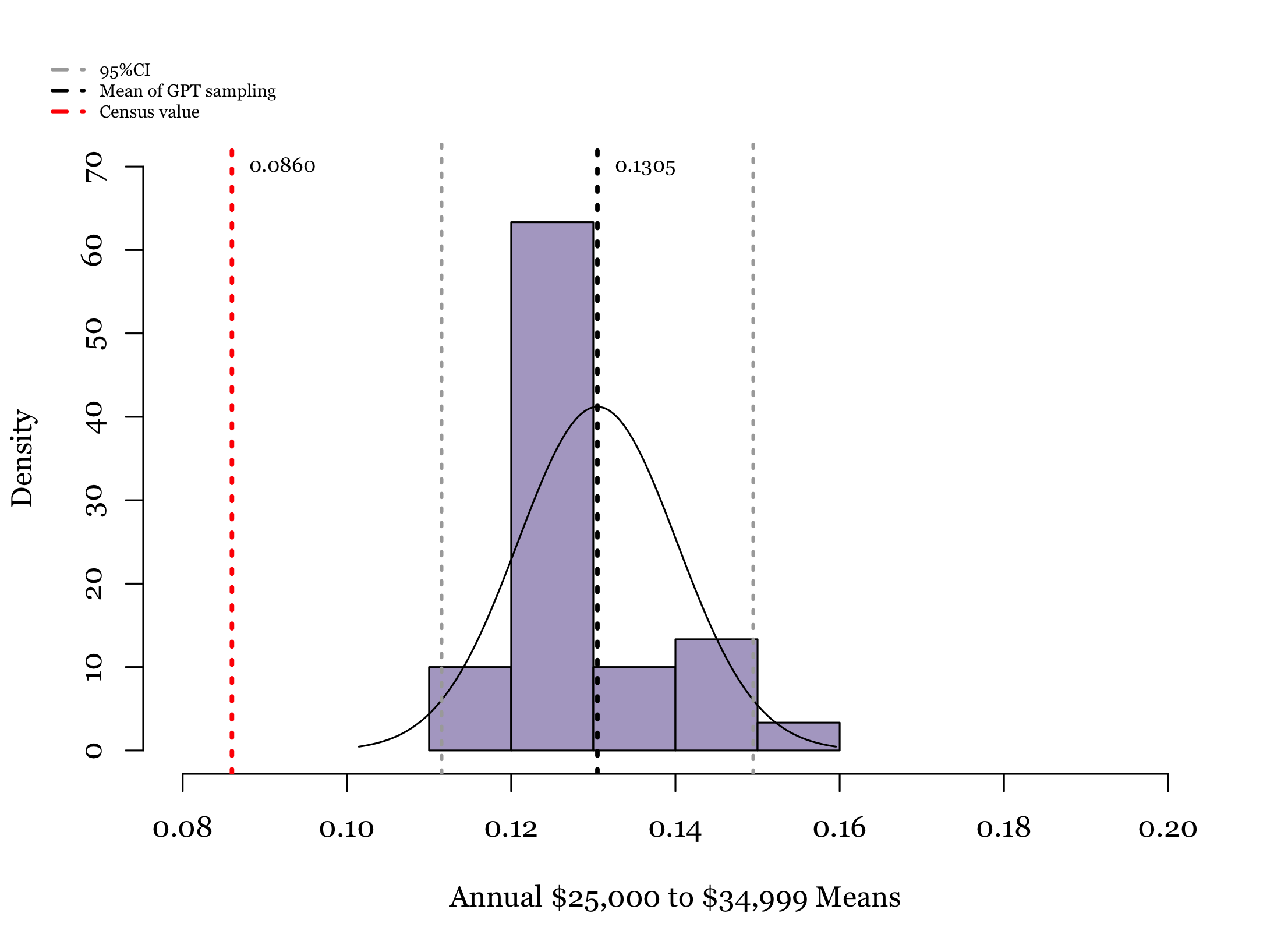}
        \caption{\$25000-\$34999}
    \end{subfigure}
    \hfill
    \begin{subfigure}{0.15\textwidth}
        \includegraphics[width=\linewidth]{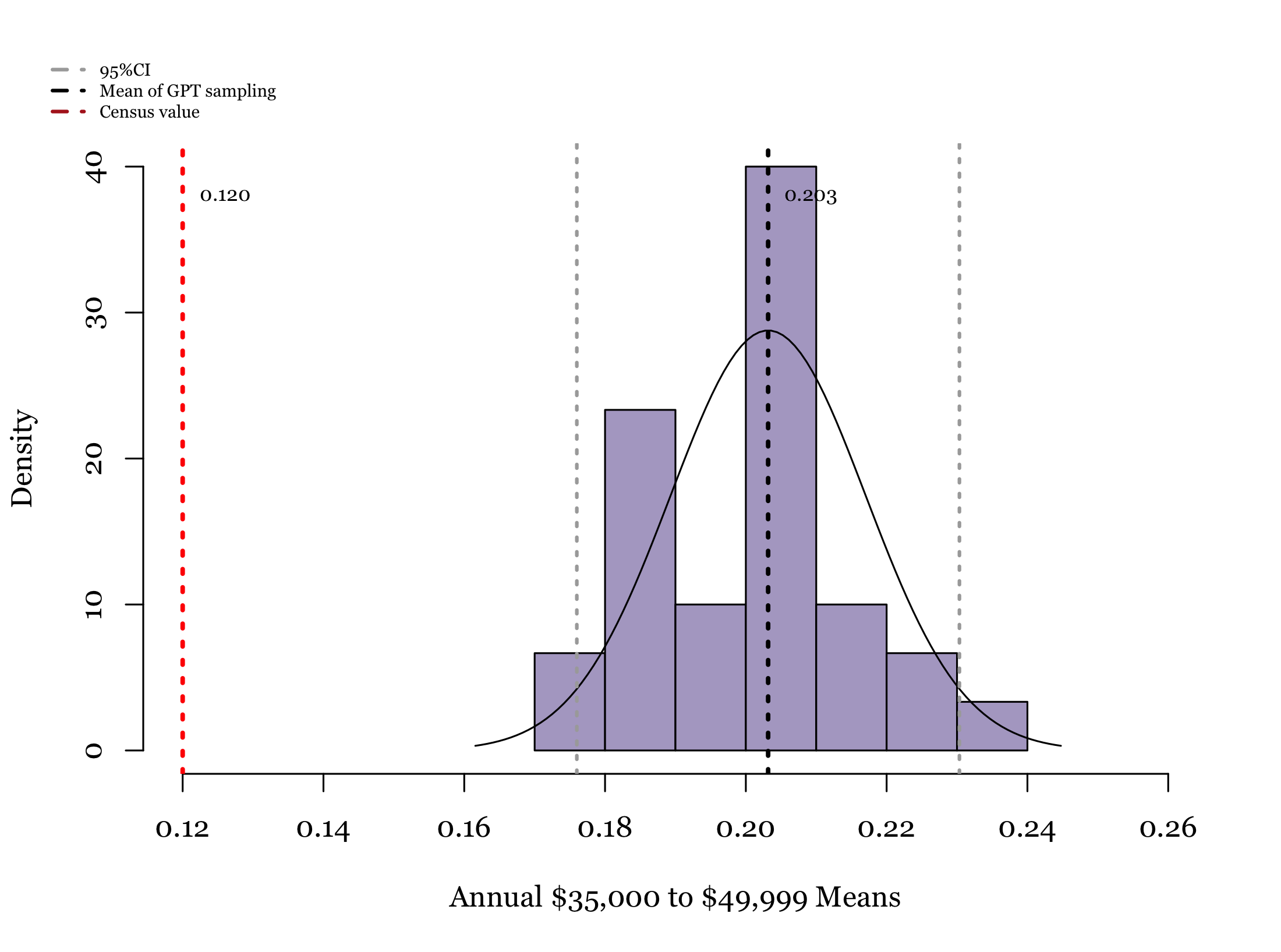}
        \caption{\$35000-\$49999}
    \end{subfigure}
    \begin{subfigure}{0.15\textwidth}
        \includegraphics[width=\linewidth]{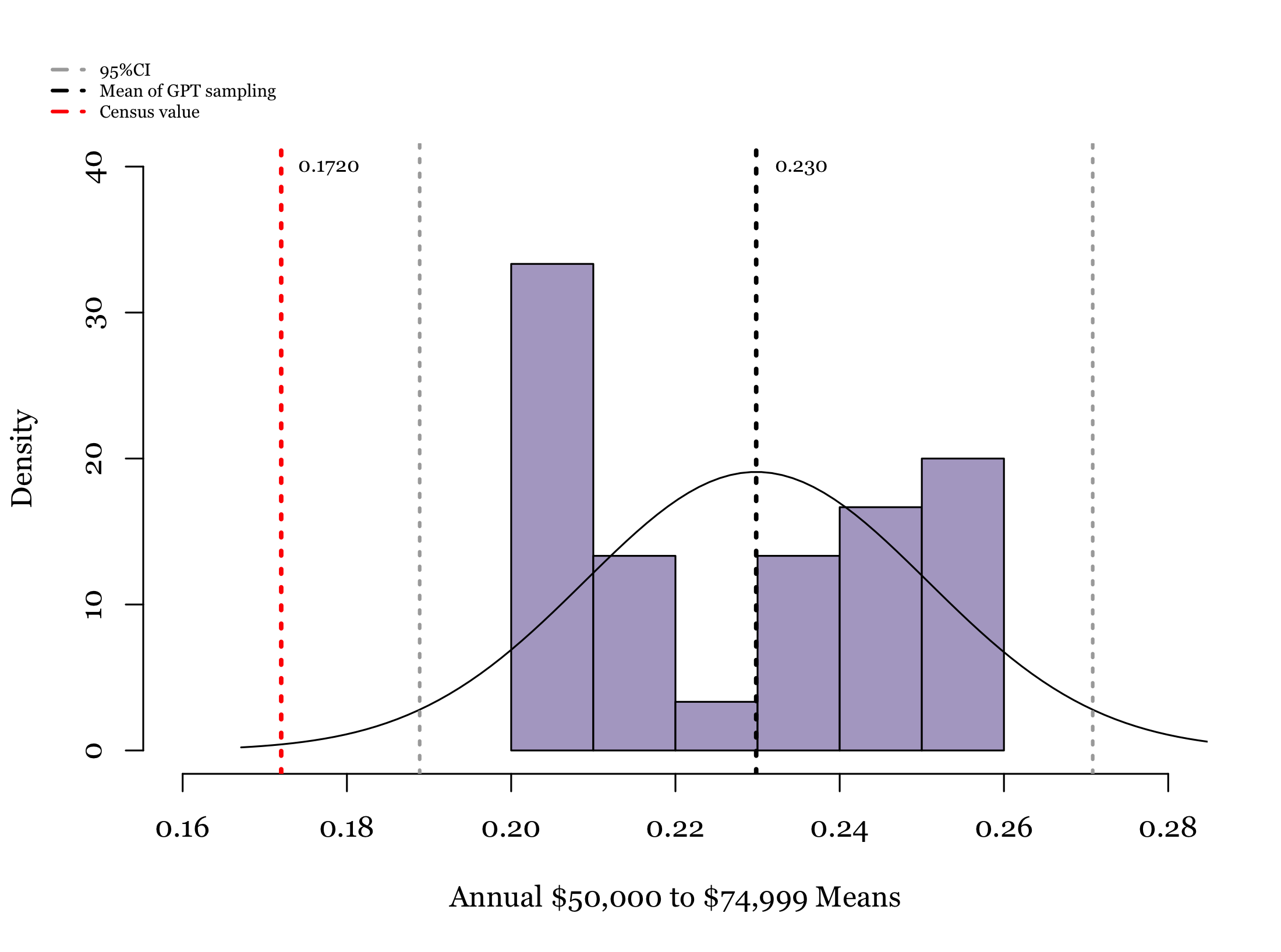}
        \caption{\$50000-\$74999}
    \end{subfigure}
    
    % Row 3
    \vspace{10pt}  
    \centering
    \begin{subfigure}{0.15\textwidth}
        \includegraphics[width=\linewidth]{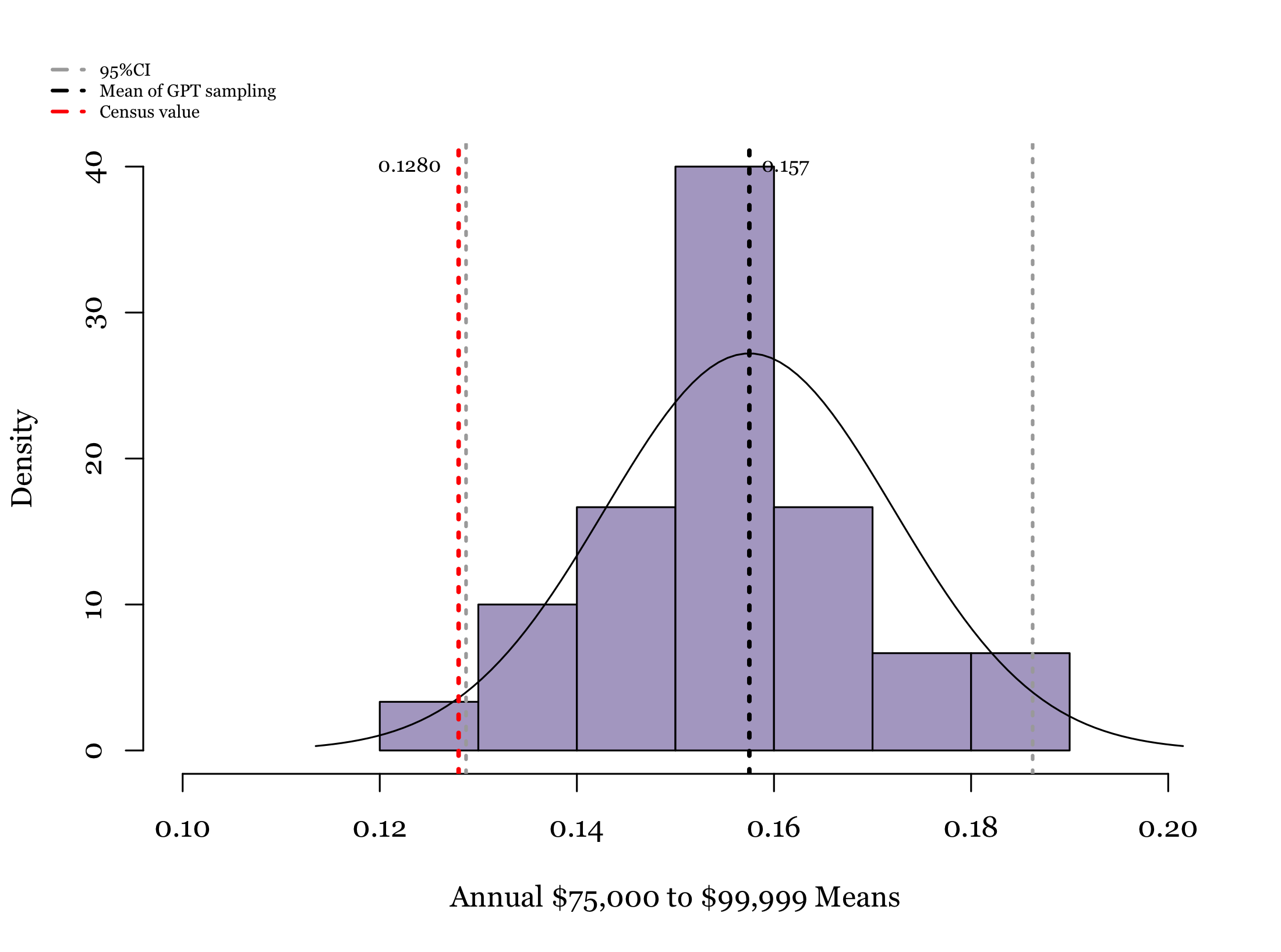}
        \caption{\$75000-\$99999}
    \end{subfigure}
    \begin{subfigure}{0.15\textwidth}
        \includegraphics[width=\linewidth]{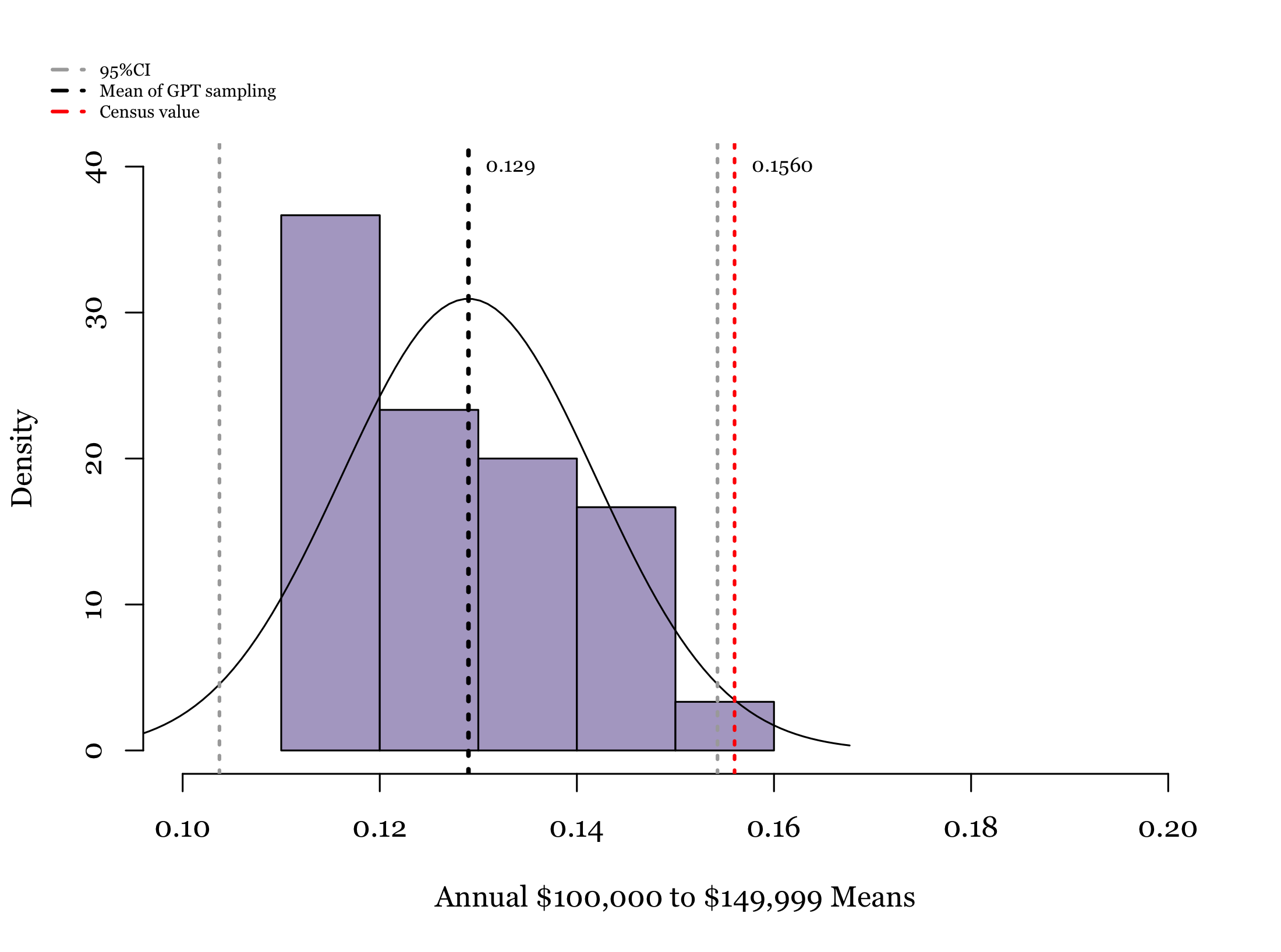}
        \caption{\$100000-149999}
    \end{subfigure}
    \hfill
    \begin{subfigure}{0.15\textwidth}
        \includegraphics[width=\linewidth]{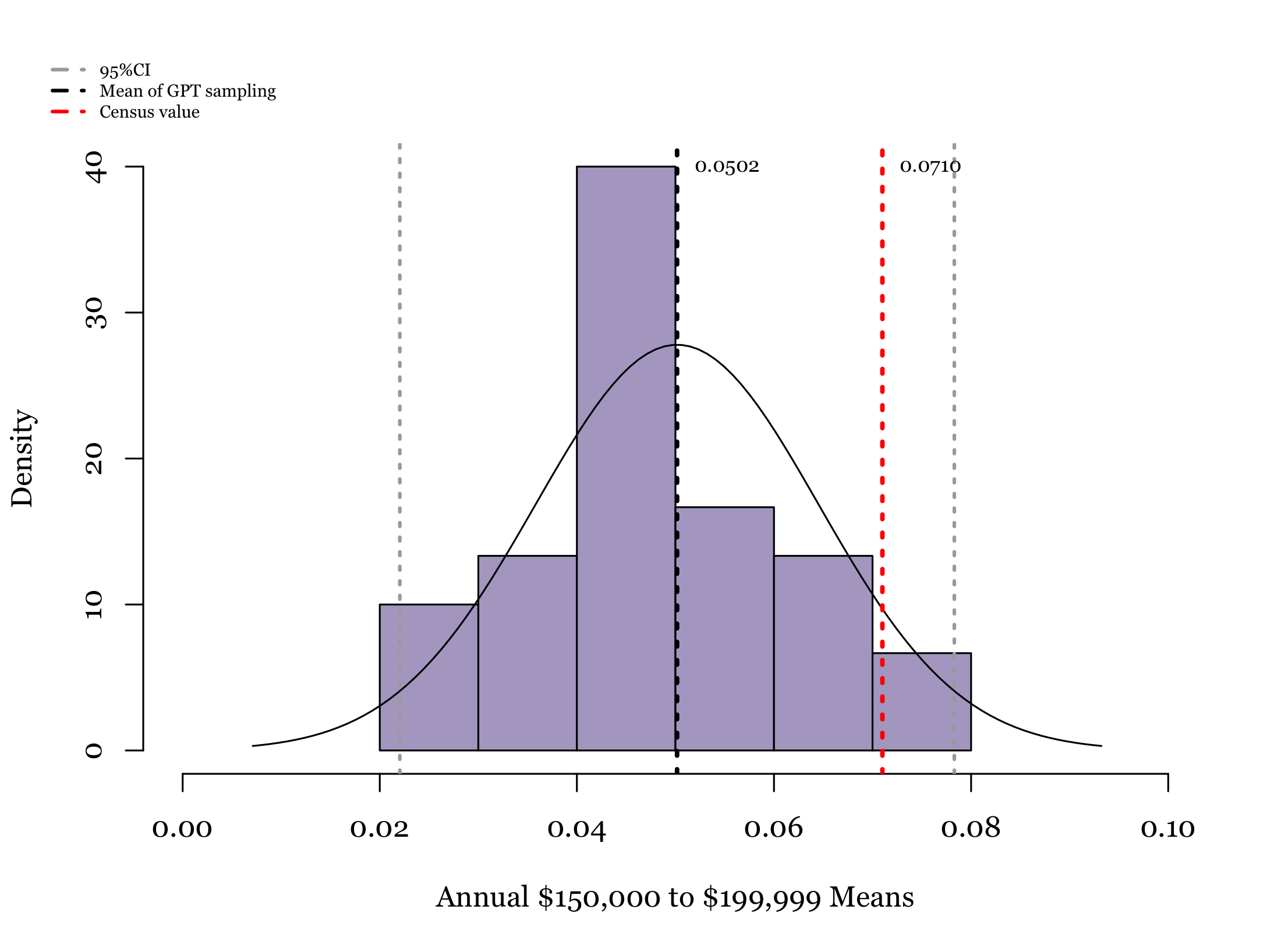}
        \caption{\$150000-\$199999}
    \end{subfigure}
    \hfill
    \begin{subfigure}{0.15\textwidth}
        \includegraphics[width=\linewidth]{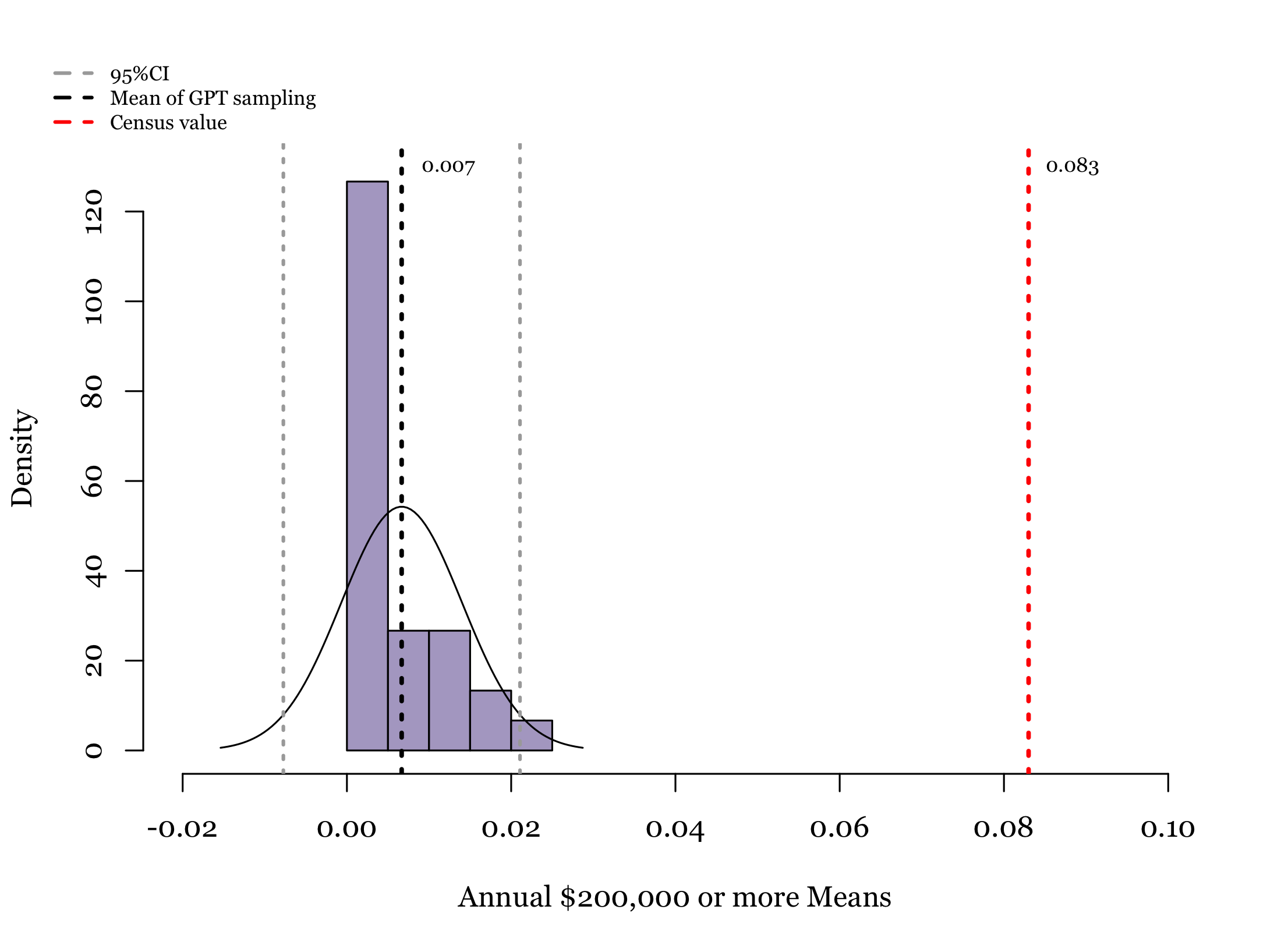}
        \caption{\$200000 or more}
    \end{subfigure}
    \caption{The sampling distribution of income groups}
    \label{f1-8}
\end{figure}
 
\textbf{6) Income:}
Figure \ref{f1-8} shows that ChatGPT underestimates the proportions of the low-income groups (less than \$10,000, \$10,000 to \$14,999) and high-income groups (\$100,000 to \$149,999, \$200,000 or more), and overestimates the proportions of the middle-income groups (\$25,000 to \$34,999, \$35,000 to \$49,999, \$50,000 to \$74,999, \$75,000 to \$99,999).

%\clearpage
\subsubsection{Population sub-group comparison}
We further investigated whether ChatGPT tends to overestimate or underestimate socioeconomic characteristics for different genders and races, focusing on educational attainment outcomes. In the 2020 US Census, the percentage of people with educational attainment beyond a bachelor's degree is 31.37\% for women and 29.34\% for men. According to the sampling distribution of the silicon population, the estimated mean proportion of women with a bachelor's degree or higher was 36\%, with the true value falling close to the confidence interval. In contrast, the proportion of men with a bachelor's degree or higher was 38.13\%, higher than the Census value. 

%According to this data, the percentage of females with educational attainment beyond a bachelor's degree is 31.37\%, while for males, it is 29.34\%. In the data generated by ChatGPT, the estimated percentage of females with educational attainment beyond a bachelor's degree is 31.14\%, with no significant difference from the true values based on chi-square test results. However, ChatGPT's data shows that the percentage of males with educational attainment beyond a bachelor's degree is 40\%, significantly higher than the actual value, indicating an overestimation by ChatGPT.

\begin{figure}[hbt!]
        % Row 1
    \centering
    \begin{subfigure}{0.23\textwidth}
        \includegraphics[width=\linewidth]{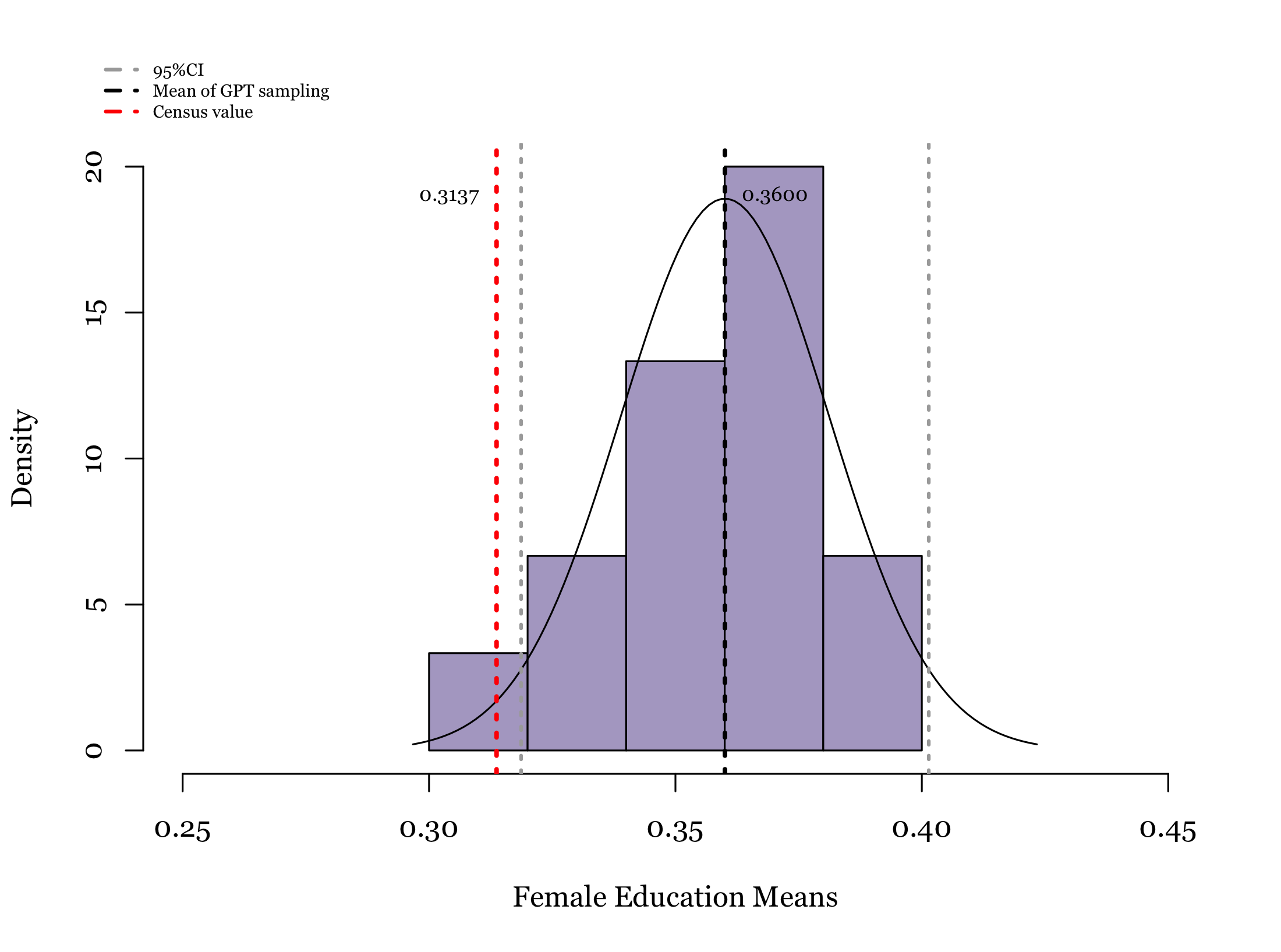}
        \caption{Women}
       \end{subfigure}
        \hfill
    \begin{subfigure}{0.23\textwidth}
        \includegraphics[width=\linewidth]{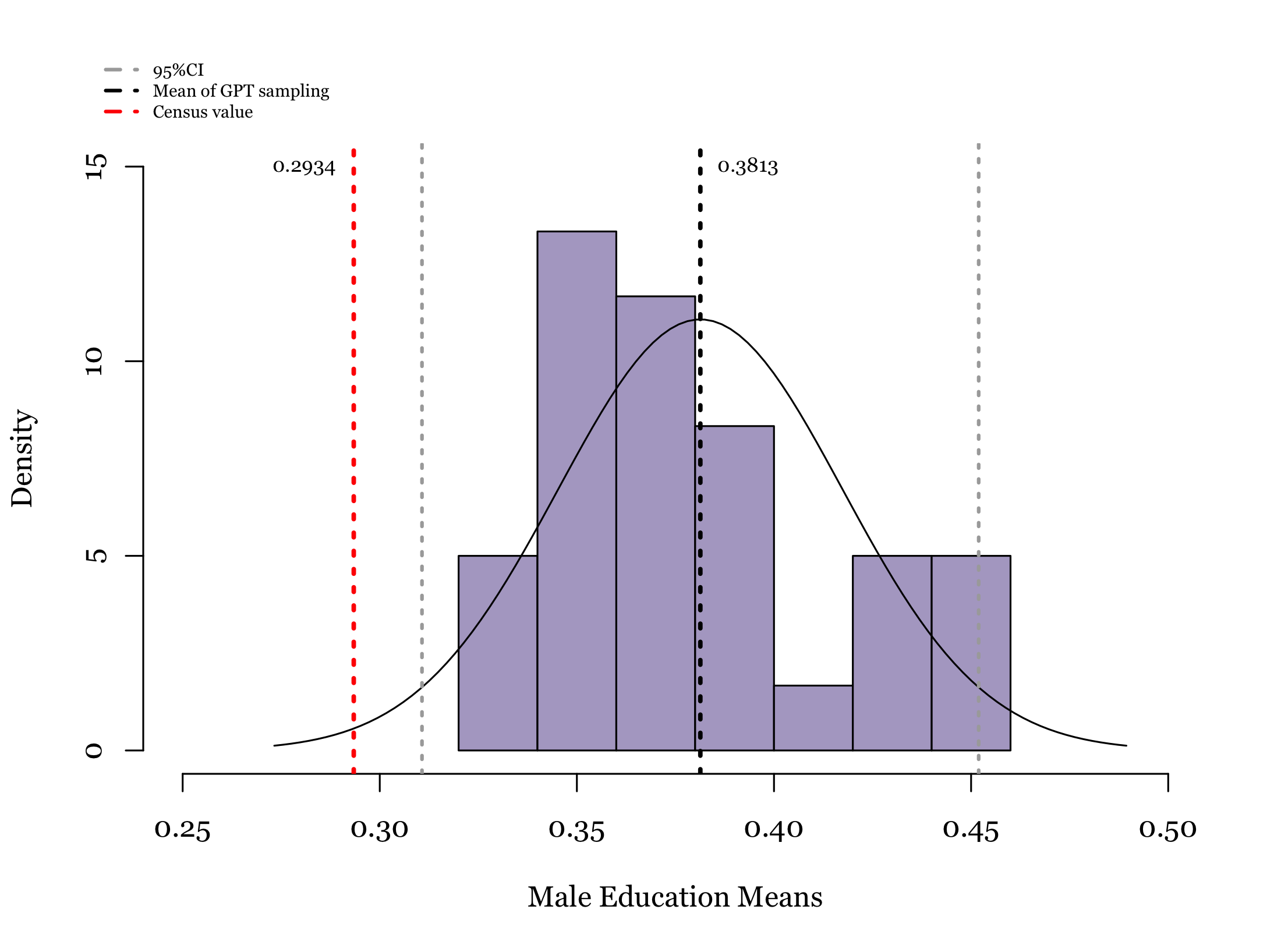}
        \caption{Men}
    \end{subfigure}
     \caption{The sampling distribution of education for gender subgroups}
    \label{f1-11}
\end{figure}

We also analyzed the educational distribution across racial groups. According to the results of the ChatGPT sampling distribution, the estimated mean proportion of individuals with a bachelor's degree or higher among Whites was 52.32\%, which is significantly higher than the Census value. For Asian, Black, and Hispanic groups, the Census values all fall within the confidence intervals, indicating a good alignment between the silicon population and the Census.

\begin{figure}[hbt!]
    \vspace{10pt}
    \centering
    \begin{subfigure}{0.23\textwidth}
        \includegraphics[width=\linewidth]{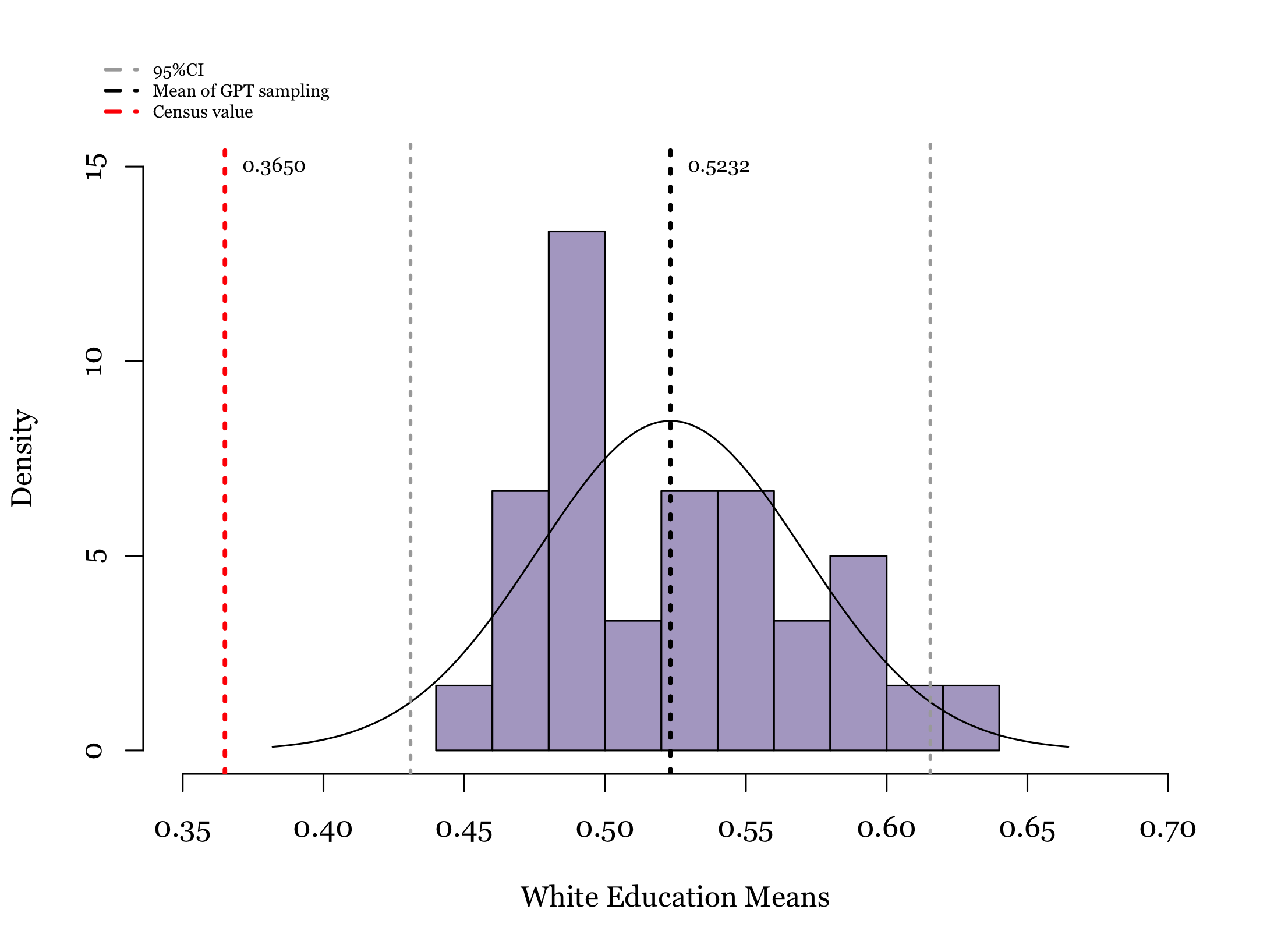}
        \caption{White}
    \end{subfigure}
    \hfill
    \begin{subfigure}{0.23\textwidth}
        \includegraphics[width=\linewidth]{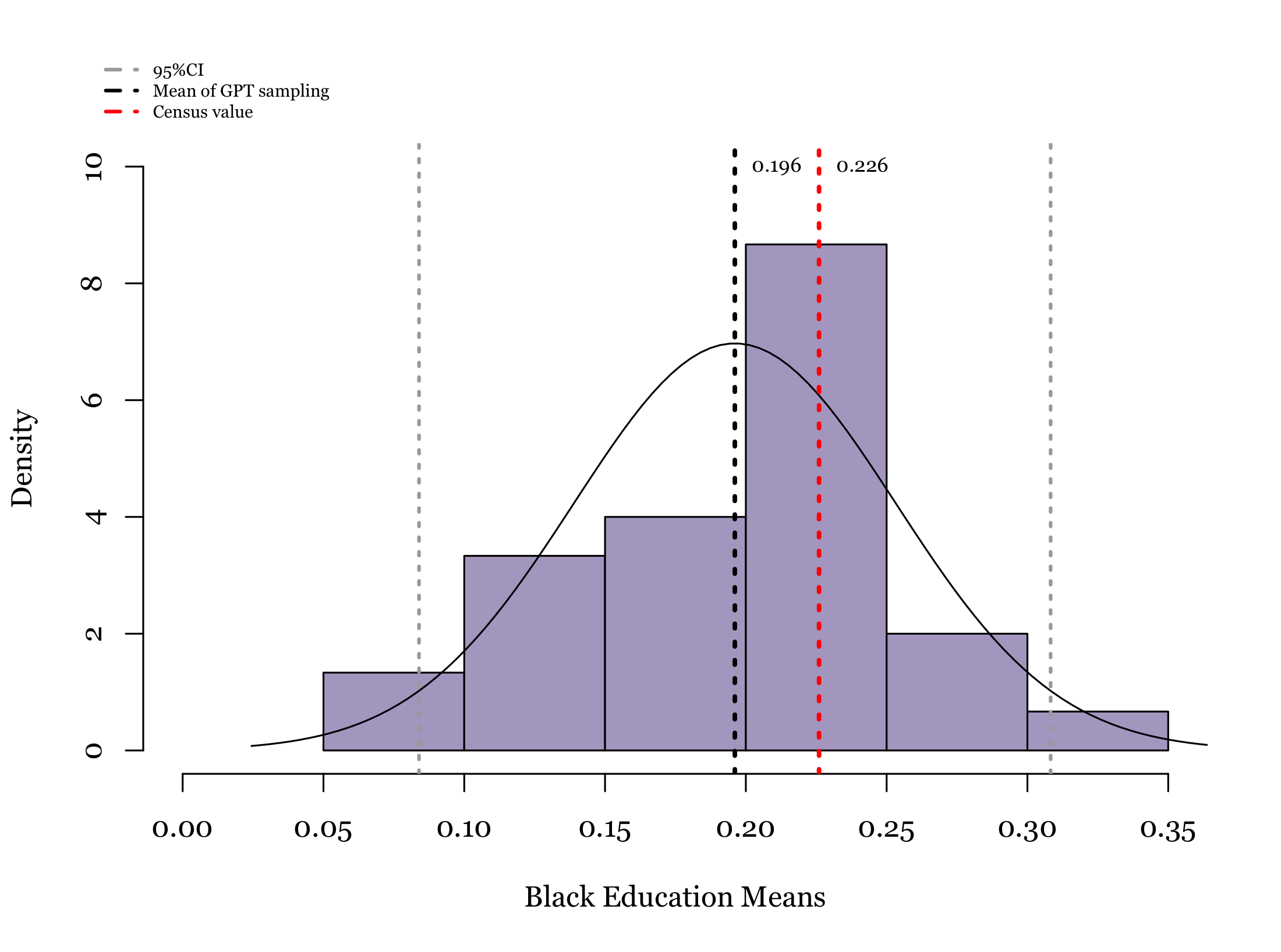}
        \caption{Black}
    \end{subfigure}  

        % Row 2
    \vspace{10pt}  
    \centering
    \begin{subfigure}{0.23\textwidth}
        \includegraphics[width=\linewidth]{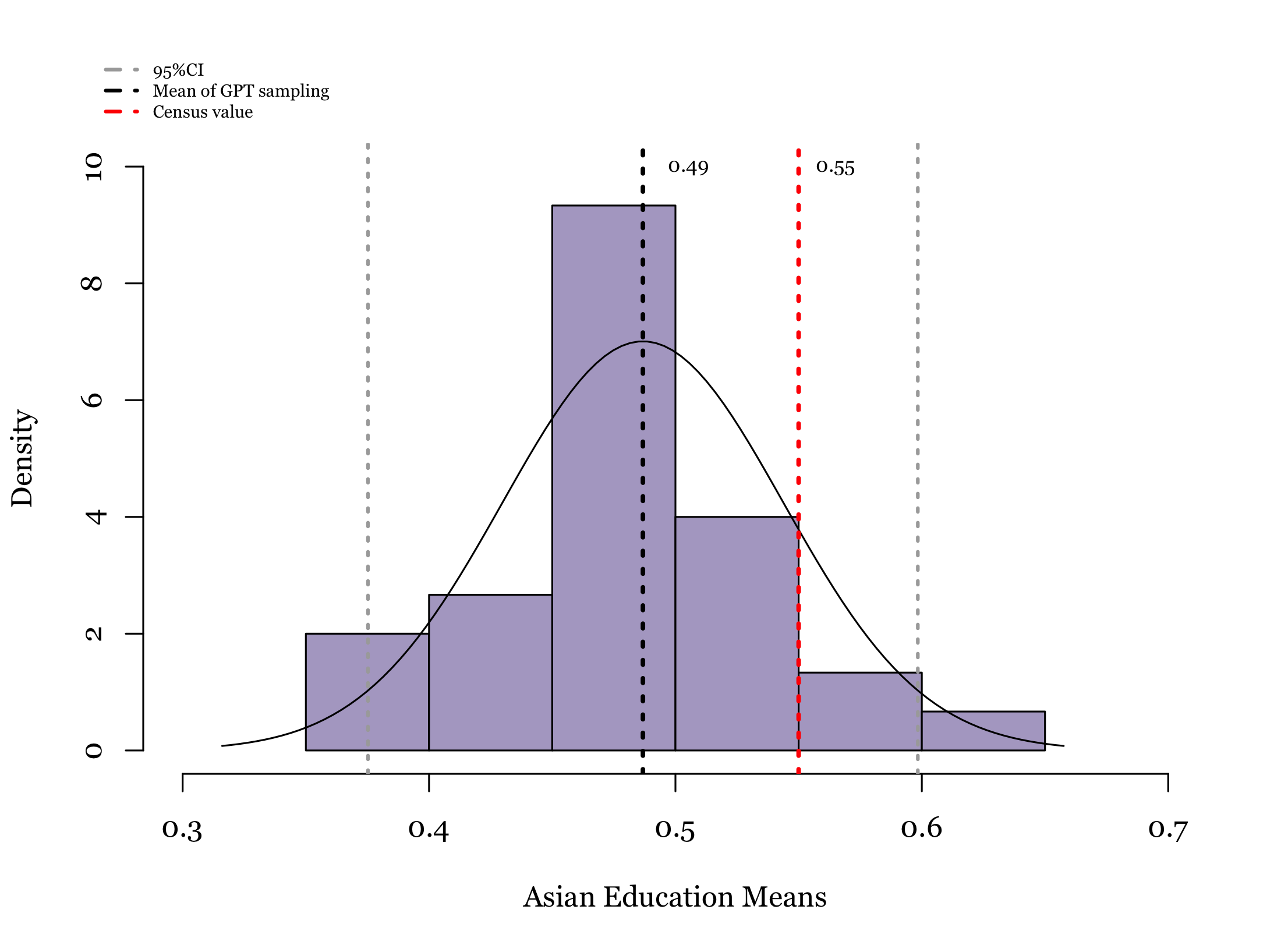}
        \caption{Asian}
    \end{subfigure}
    \hfill
    \begin{subfigure}{0.23\textwidth}
        \includegraphics[width=\linewidth]{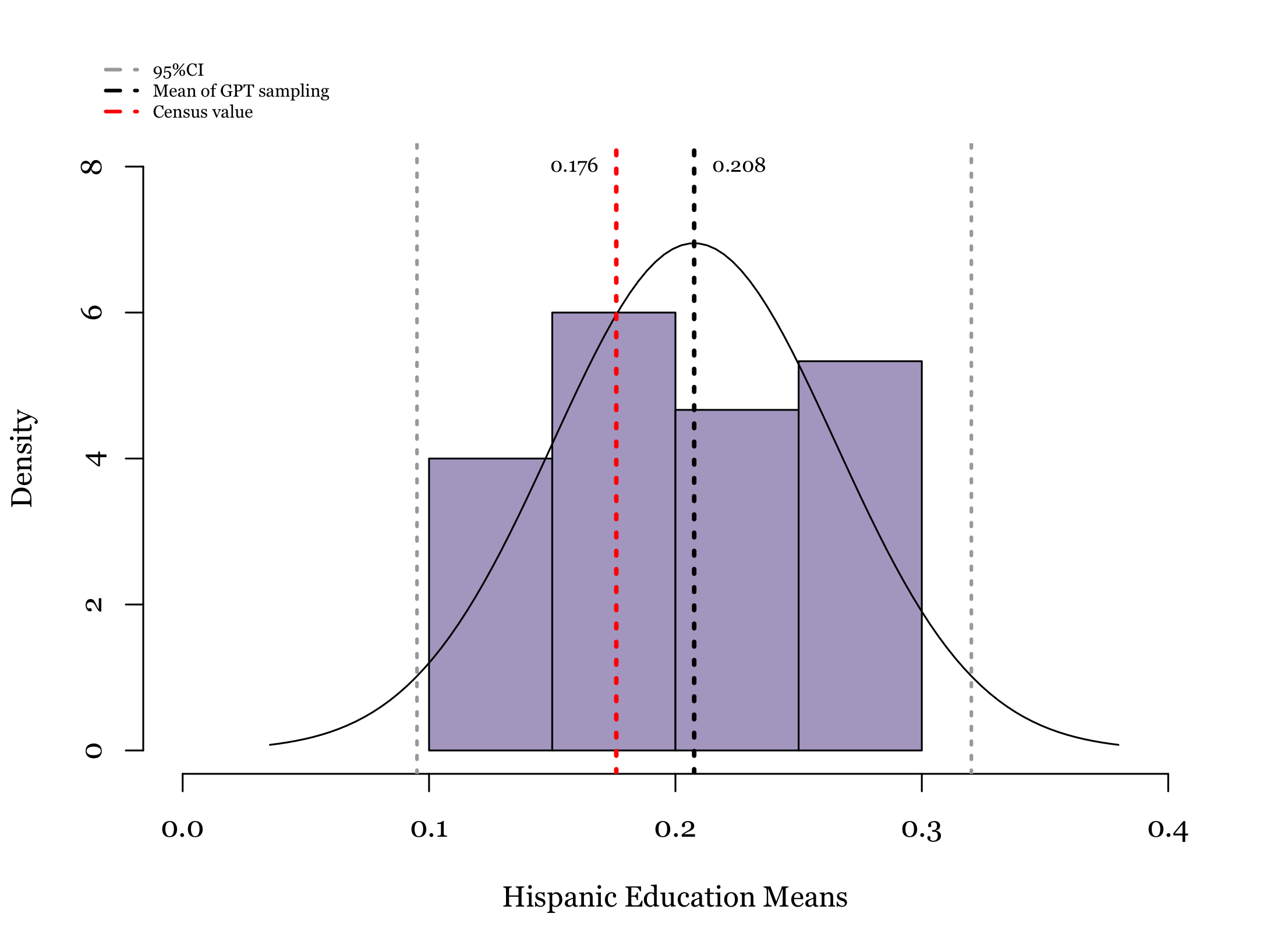}
        \caption{Hispanic}
    \end{subfigure}
    \caption{The sampling distribution of education for racial subgroups}
    \label{f1-12}
\end{figure}

%\begin{table*}[ht] {\columnwidth}
%\caption{Distribution of educational attainment between genders and ethnic groups}
%\label{S1-4}
%\centering
%\begin{tabular}{ccccc}
%\hline
% & \textbf{Bachelor+ size}& \textbf{Total sample size}& \textbf{\% GPT}& \textbf{\% Census}\\
%\hline
%\textbf{Males} & 200 & 499 & 40.08\% & 29.34\% \\
%\hline
%\textbf{Females} & 156 & 501 & 31.14\% & 31.37\% \\
%\hline
%\textbf{Whites} & 207 & 409 & 50.61\% & 36.50\% \\
%\hline
%\textbf{Asians } & 66 & 149 & 44.30\% & 55.00\% \\
%\hline
%\textbf{Blacks} & 38 & 212 & 17.92\% & 22.6\% \\
%\hline
%\end{tabular}
%\end{table*}

The above results suggest that ChatGPT's bias seems to manifest as an overestimation of advantaged groups rather than an underestimation of disadvantaged ones. This pattern aligns with recent research by Bloomberg on biases in ChatGPT recruitment. The study indicates that while recruitment algorithms may not explicitly show preferences for specific demographics, they can still influence outcomes by favoring certain criteria \cite{bloomberg2024}. Algorithms that appear unbiased may conceal their favoritism toward particular demographics.

\subsection{Knowledge and Performance Gap}
The current GPT model may have already learned the distribution of socioeconomic characteristics of the US 2020 population from publicly available information. We directly inquired ten times in a conversational style about the distributions of these six characteristics via GPT-3.5-turbo. The responses from GPT indicate that it has indeed acquired this information. For example, it consistently reported that the proportion of women was 51\% and the proportion of Black individuals was between 12.1\% and 12.4\%.

However, as noted above, when we asked GPT to randomly select a respondent multiple times from the US 2020 population, it failed to create a sample in which the proportion of Black individuals matched the corresponding Census figures. This discrepancy highlights a disconnect between what GPT knows and its ability to implement that knowledge through random sampling.

 \section{STUDY 2: SILICON POPULATION INCOME AND GENDER ATTITUDES}
 
\subsection{Attitudinal Questions}
We selected questions regarding income inequality and redistribution and questions related to gender inequality from the WVS dataset (\hyperref[appendix3]{Appendix: WVS Questions}). For the income-related attitudes (score ranging from 1 to 10), a higher score indicates a stronger belief in meritocracy, suggesting that personal effort should be rewarded and is a key income determinant. For the four gender role-related attitudes (score ranging from 1 to 4), a higher score reflects a higher level of disagreement with those statements about traditional gender roles.

\subsection{Experiment setting}
%We used the GPT-3.5-turbo-0613 version of GPT-3.5-turbo to generate answers to these questions. 
We tasked GPT-3.5-turbo-0613 to act as a virtual respondent with particular demographic characteristics answering attitudinal questions from the WVS dataset. The prompts we used were in a questionnare format (prompt in Github\footnote{https://anonymous.4open.science/r/Surrogate/README.md}. 

If the virtual respondent provides an unclear, uncertain, or no answer, it will be prompted again until it responds accurately. For income inequality questions, all responses were collected after two rounds of inquiries, while for gender inequality questions, five rounds of inquiries were needed to obtain all eligible answers. In addition, similar to Study 1, we generated a sample of 200 data points and repeat this process 30 times to form sampling distributions of these income attitude scores and gender attitude scores. The mean values of these sampling distributions can be compared with the sample mean of the WVS sample to evaluate how closely responses from silicon samples align with those from human respondents.

\subsection{Results}

\subsubsection{Income Inequality and Redistribution}

\begin{figure}[hbt!]
    \hspace*{\fill}
    \centering
     \begin{subfigure}{0.23\textwidth}
        \includegraphics[width=\linewidth]{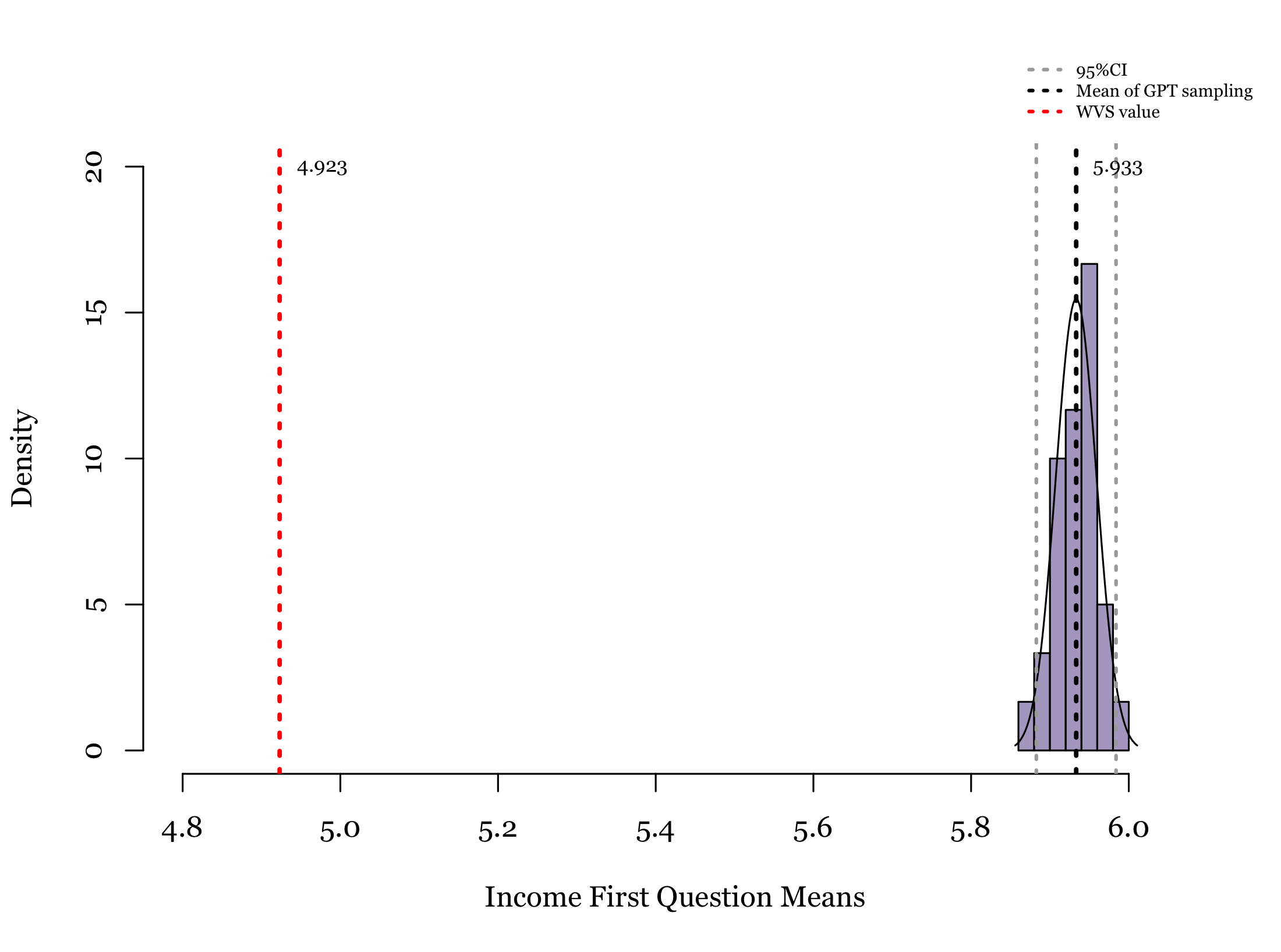}
        \caption{Q106: More equal income vs. More individual efforts}
    \end{subfigure}  
       \hfill
    \begin{subfigure}{0.23\textwidth}
        \includegraphics[width=\linewidth]
        {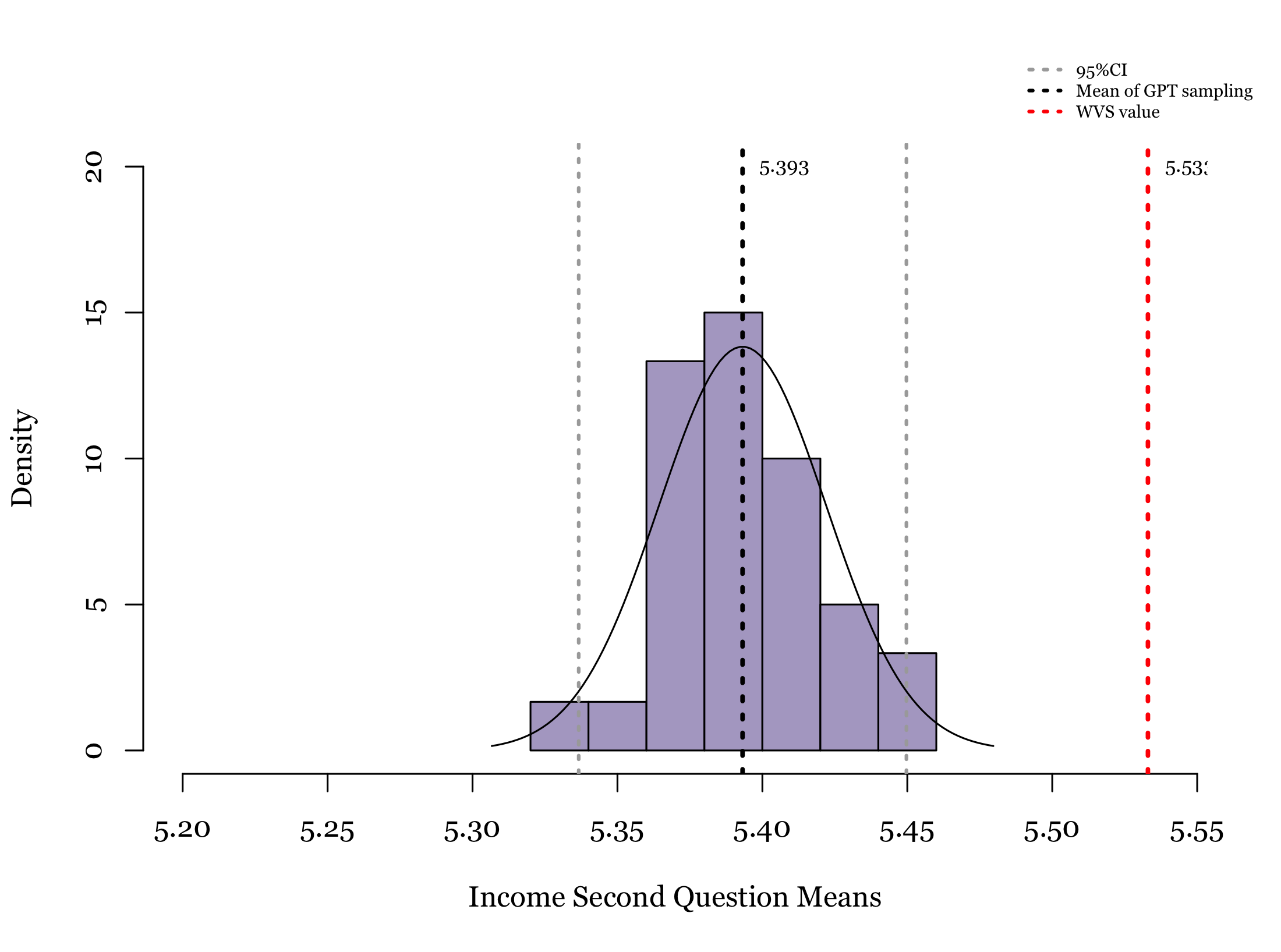}
        \caption{Q108: More government vs. More personal responsibility}
    \end{subfigure}  
    \caption{Sampling distribution of income-related attitudes; Red line - WVS}
    \label{s2-f1}
       \end{figure}

Table \Ref{tab:AR-t1} presents the means and standard deviations of the scores for the two income-related questions from the WVS and one sample of the GPT-3.5-0613. First, for both questions, the s.d. of the scores from the GPT agent sample was substantially smaller than that of the WVS, suggesting less variability in the GPT responses. The sampling distribution (Figure \ref{s2-f1}) of the multiple GPT samples suggests that GPT agents scored higher than WVS respondents in Q106, indicating GPT's greater inclination towards meritocracy. GPT agents scored lower in Q108, implying slightly more support for government intervention. These findings do not show that GPT agents demonstrate a consistent inclination toward certain income inequality views.

% income mean
\begin{table}[htp]
\centering
\caption{Sample distribution of income-related attitudinal scores}
\label{s1-sampledist}
\resizebox{0.5\textwidth}{!}{
    \begin{tabular}{cccccccc}
    \hline
         &  &  \multicolumn{2}{c}{WVS}&  \multicolumn{2}{c}{GPT-turbo-0613} &  \multicolumn{2}{c}{t-test: WVS vs GPT}\\
         &  N &  Mean&  Std. dev.&  Mean&  Std. dev. & Diff & t-score\\
         \hline
         Q106&  2,463&  4.936&  2.840&  5.831&  1.246&  -0.894*** & -14.315\\
         Q108&  2,463&  5.557&  2.969&  5.420&  1.369&  0.137** & 2.084\\
         \hline
    \end{tabular}}
    \begin{tablenotes}
      \small
      \item Notes: * p ~\textless 0.05, ** p ~\textless 0.01, *** p ~\textless 0.001
    \end{tablenotes}
    \label{tab:AR-t1}
\end{table}

\begin{figure}[hbt!]
    \hspace*{\fill}
     \begin{subfigure}{0.23\textwidth}
        \includegraphics[width=\linewidth]
        {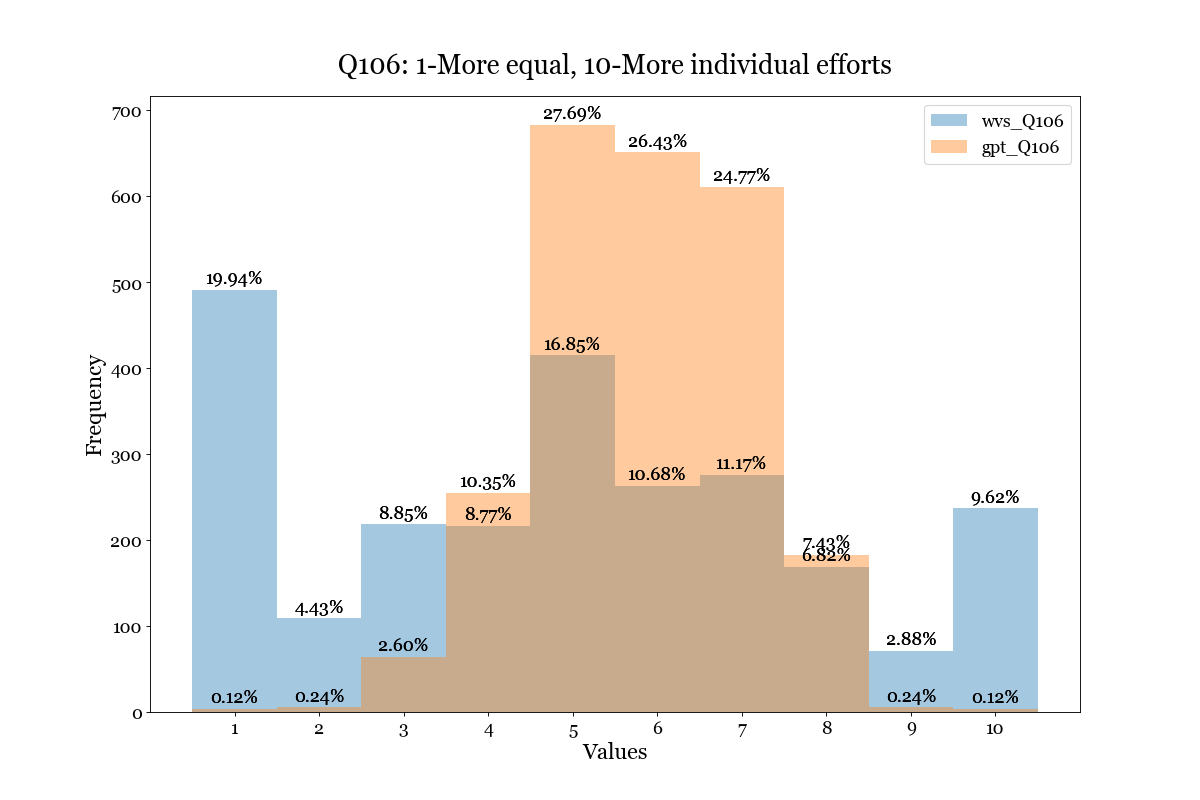}
        \caption{Q106: More equal income vs. More individual efforts}
    \end{subfigure}  
       \hfill
    \begin{subfigure}{0.23\textwidth}
        \includegraphics[width=\linewidth]
        {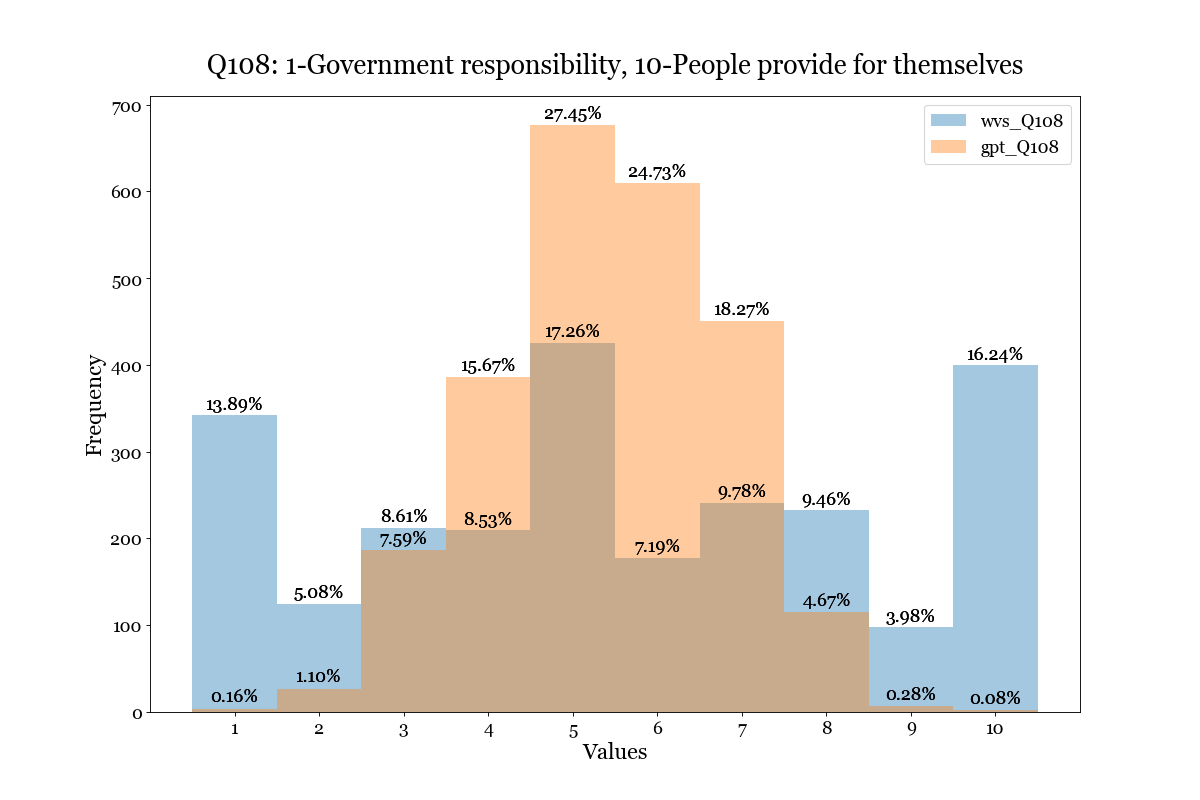}
        \caption{Q108: More government vs. More personal responsibility}
    \end{subfigure}  
    \caption{Sample distribution of income attitude scores}
    \label{s2-income}
       \end{figure}

Figure \ref{s2-income} illustrates the distribution of scores from one sample of the GPT agents and WVS responses. The most notable difference is the normal distribution of the responses from the GPT agents. In contrast, the answers from WVS respondents tend to cluster around the middle and at both extremes, indicating more polarized views on these statements. As noted earlier, the variability in responses from the GPT agents is significantly smaller than that of human respondents. 

% income

\subsubsection{Gender Roles and Gender Inequality}
% gender
Table \Ref{tab:AR-t2} presents the means and standard deviations of gender-related attitudes from both GPT and WVS samples. Again, the scores from GPT agents exhibit much smaller variations compared to those of human respondents.

\begin{figure}[hbt!]
 %row1
    \hspace*{\fill}
    \centering
     \begin{subfigure}{0.23\textwidth}
        \includegraphics[width=\linewidth]{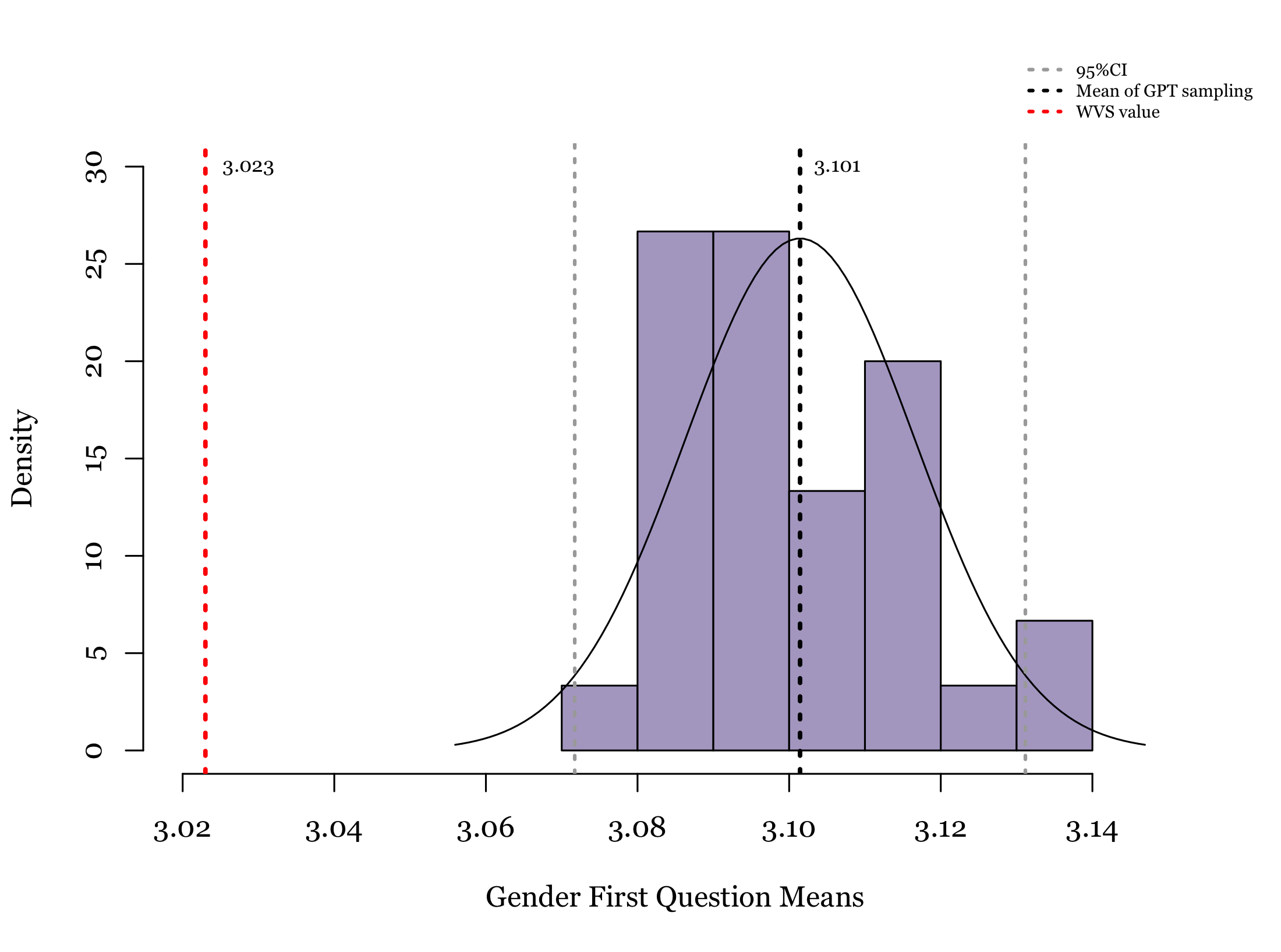}
        \caption{Q28: Mother work, children suffer}
    \end{subfigure}  
       \hfill
    \begin{subfigure}{0.23\textwidth}
        \includegraphics[width=\linewidth]
        {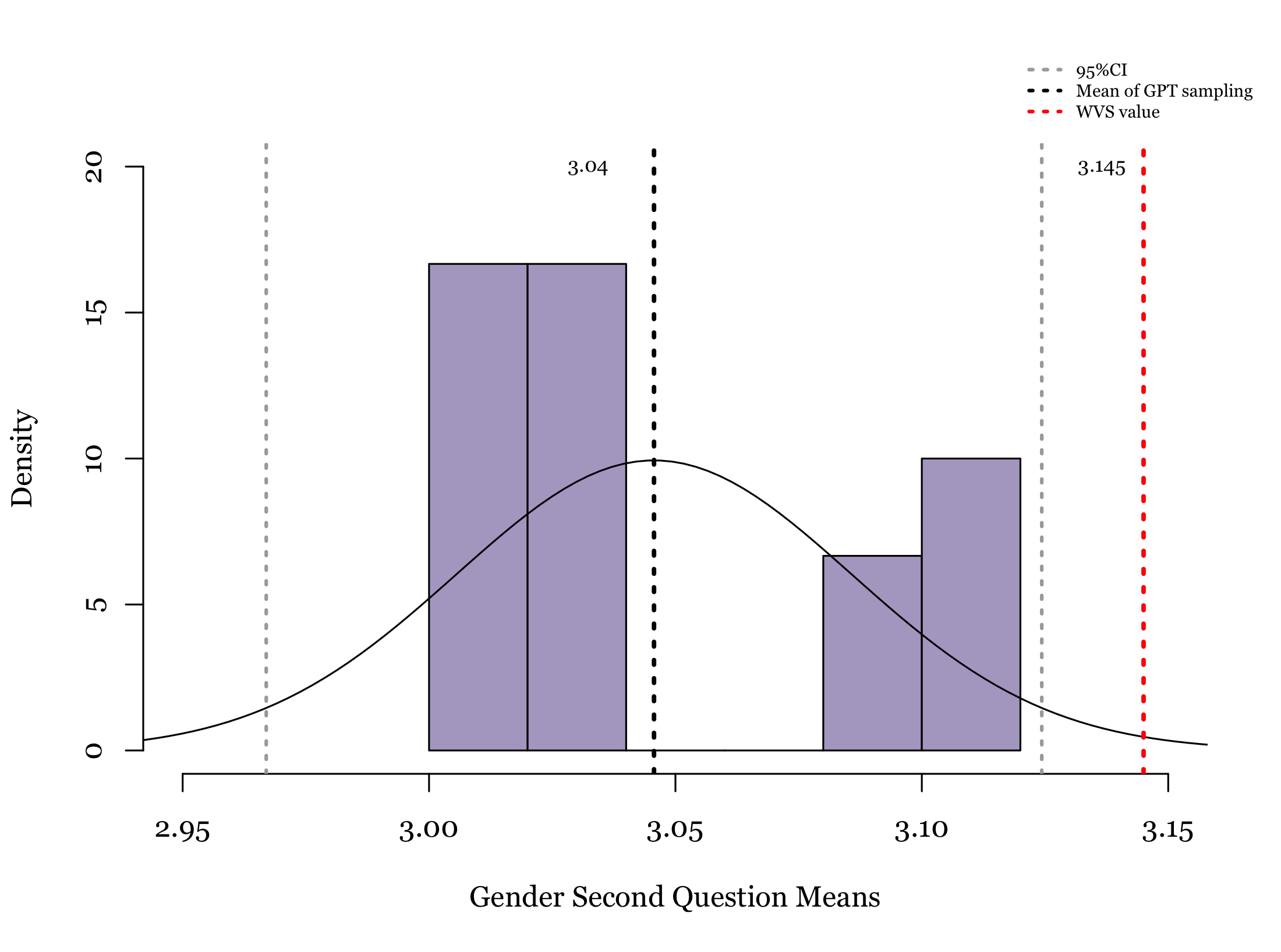}
        \caption{Q29: Men better political leaders}
    \end{subfigure}  
    
    %row2
      \vspace{10pt}    
          \hfill
        \begin{subfigure}{0.23\textwidth}
        \includegraphics[width=\linewidth]
        {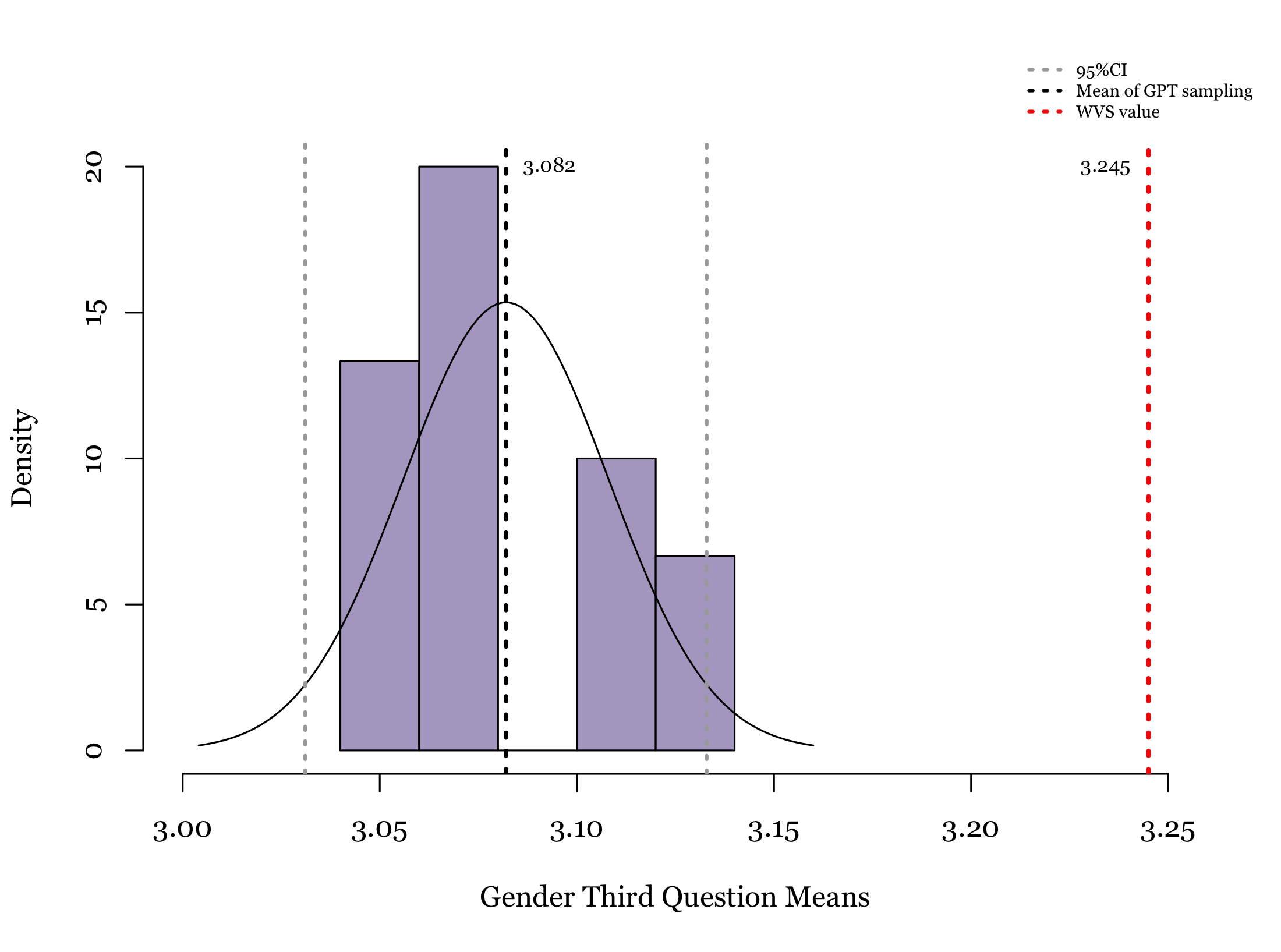}
        \caption{Q31: Men better business executives}
    \end{subfigure}
        \begin{subfigure}{0.23\textwidth}
        \includegraphics[width=\linewidth]
        {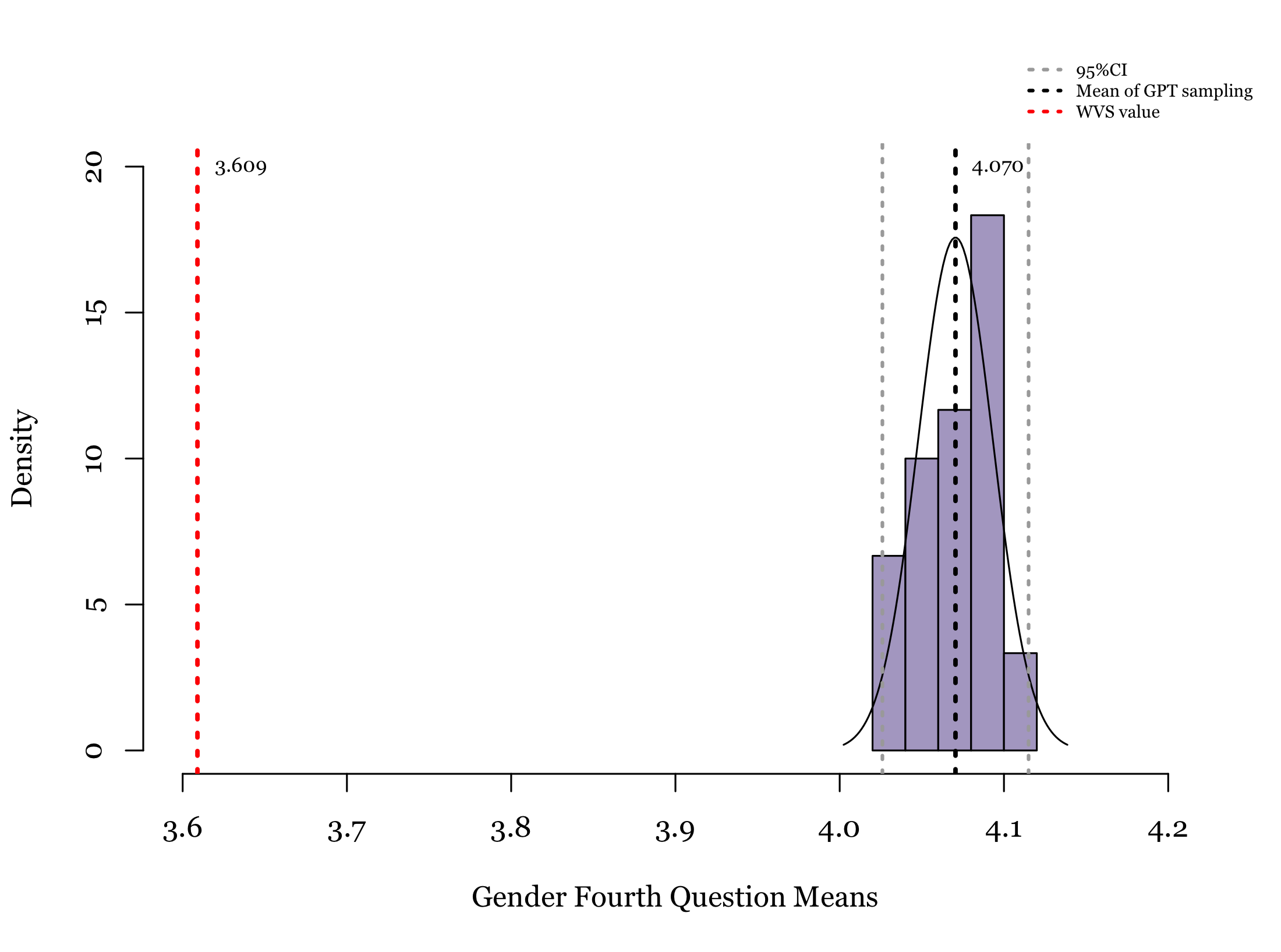}
        \caption{Q35: Women more money more trouble}
    \end{subfigure}    
    \caption{Sampling distribution of gender attitudes; Red line - WVS}
    \label{s2-gf1}
       \end{figure}

Sampling distribution results from Figure \Ref{s2-gf1} indicate significant differences, particularly for Q35, the final question regarding women out-earning men, where GPT agents scored much higher. Higher scores reflect a stronger disagreement with traditional gender roles; therefore, if GPT consistently scores higher than WVS respondents, we could conclude that GPT agents are more gender egalitarian. However, we do not find consistent evidence of a gender-egalitarian ideology embedded in GPT's world, contrary to findings from previous studies~\cite{rutinowski2024self}.

\begin{table}[htp]
    \centering
    \caption{Sample distribution of gender-related attitudinal scores}
\resizebox{0.5\textwidth}{!}{
    \begin{tabular}{cccccccc}
    \hline
         &  &  \multicolumn{2}{c}{WVS}&  \multicolumn{2}{c}{GPT-turbo-0613}&  \multicolumn{2}{c}{t-test: WVS vs GPT}\\
         &  N &  Mean&  Std. dev.&  Mean&  Std. dev.& Diff & t-score\\
         \hline
         Q28&  2437&  3.022&  0.728&  3.133&  0.432& -0.111***&-6.480\\
         Q29&  2437&  3.145&  0.751&  3.104&  0.405& 0.041**&2.374\\
         Q31&  2437&  3.248&  0.709&  3.089&  0.328& 0.160***&10.090\\
         Q35&  2437&  3.612&  0.633&  4.051&  0.580& -0.439***&-25.232\\
         \hline
    \end{tabular}}
     \begin{tablenotes}
      \small
      \item Notes: * p ~\textless 0.05, ** p ~\textless 0.01, *** p ~\textless 0.001
    \end{tablenotes}
    \label{tab:AR-t2}
\end{table}

\begin{figure}[hbt!]
    \hspace*{\fill}
     \begin{subfigure}{0.23\textwidth}
        \includegraphics[width=\linewidth]
        {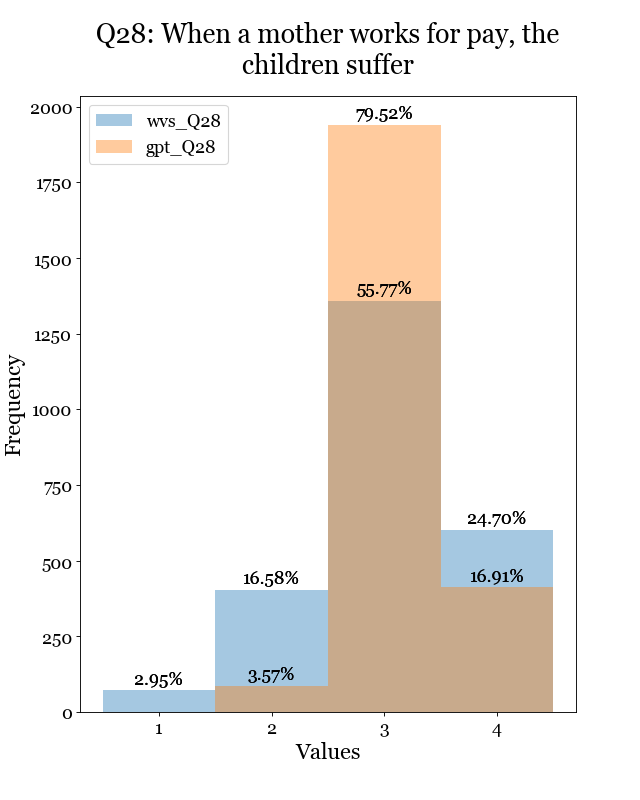}
        \caption{Q28: Mother work, children suffer}
    \end{subfigure}  
       \hfill
    \begin{subfigure}{0.23\textwidth}
        \includegraphics[width=\linewidth]
        {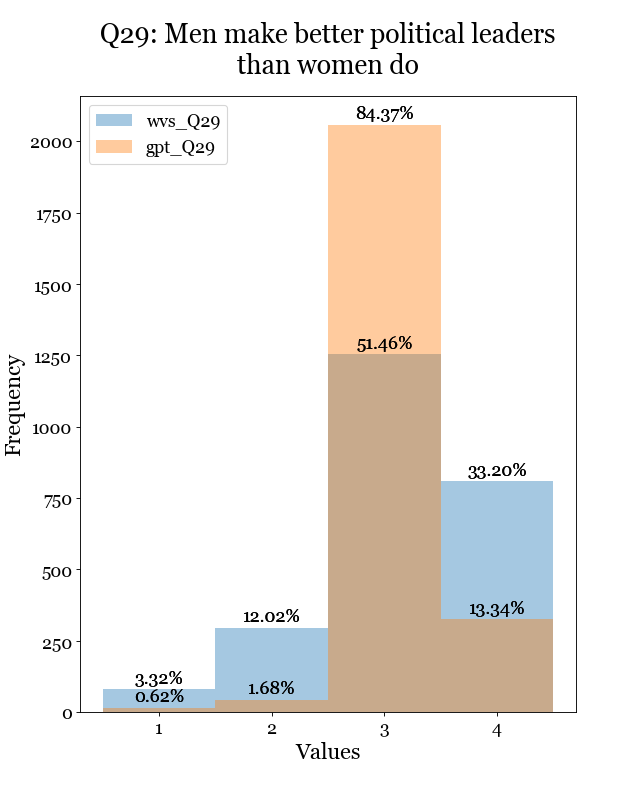}
        \caption{Q29: Men better political leaders}
    \end{subfigure} 

       %row2
      \vspace{10pt}    
     \hfill
    \begin{subfigure}{0.23\textwidth}
        \includegraphics[width=\linewidth]
        {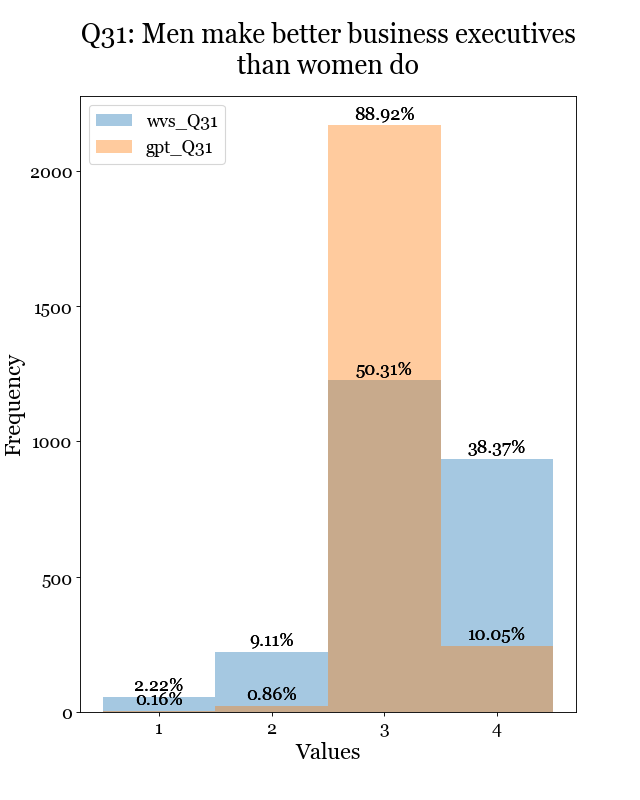}
        \caption{Q31: Men better business executives}
    \end{subfigure}  
     \hfill
    \begin{subfigure}{0.23\textwidth}
        \includegraphics[width=\linewidth]
        {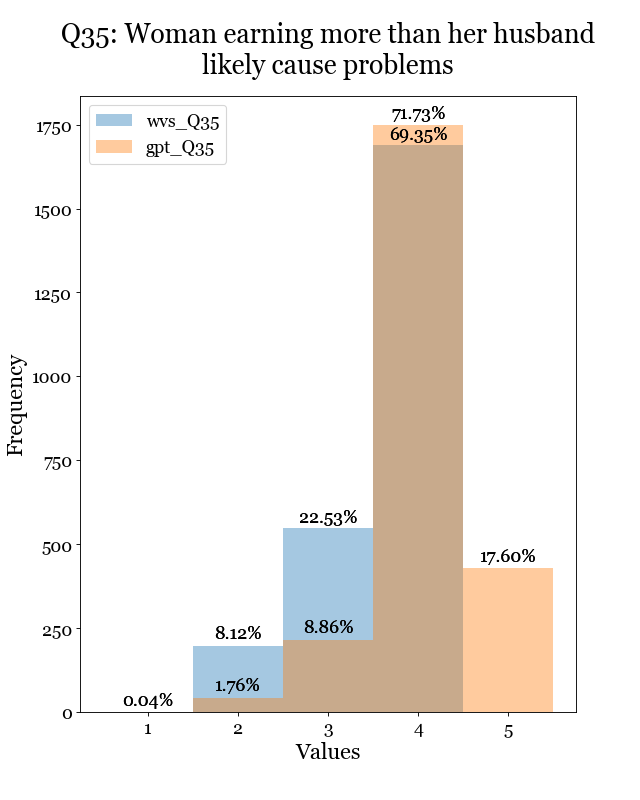}
        \caption{Q35: Women more money more trouble}
    \end{subfigure}  
    \caption{Sample distribution of gender attitude scores}
    \label{s2-gendersample}
       \end{figure}

Figure \ref{s2-gendersample} illustrates the score distributions from silicon and WVS samples. For all four questions, GPT agents accurately identify the score category with the highest percentage of responses from WVS. However, nearly all GPT responses are concentrated in this most frequent category, resulting in significantly less variation compared to human respondents. 

%This response pattern with much lower entropy is observed across various gender, racial, and educational groups.

%The scores given by GPT-0613 respondents for gender inequality questions are also more concentrated compared to WVS. 

%However, unlike income inequality, the score distribution for gender inequality questions among GPT-0613 respondents is more concentrated than that of GPT respondents based on GPT-3.5-turbo. Additionally, for gender inequality question 2, the score distribution among GPT-0613 respondents no longer shows high consistency with WVS. 

 \section{CONCLUSION}

The mixed conclusions from earlier studies that focus extensively on the alignment of social or political attitudes between LLM agents and humans call for a comprehensive re-evaluation of LLM's abilities to generate human-like responses in social surveys. Different from many earlier studies that extensively focus on the level of accuracy in the prediction generated by language agents, this paper focuses on both the point estimation of these responses as well as the distribution of these responses. We proposed the application of the \textit{Central Limit Theorem} to capture the parameter of the silicon population and compare it with human benchmark values. Overall, we aim to illuminate the understanding of the human population in the eyes of LLMs. 
%Moreover, instead of only focusing on the sample ``point estimation'', we also examine the distribution of these responses. 

Our findings reveal instances of commendable alignment with the US 2020 population, alongside more evident biases, as well as significant deviations from human self-reported responses to attitudinal inquiries. Utilizing the repeated sampling method offers significant advantages in constructing a sampling distribution that allows for the identification of the GPT- or silicon-population parameter and facilitates comparison with Census or survey values. We find that ChatGPT only slightly underestimates the proportion of women and correctly estimates the mean age of the US 2020 population. However, GPT cannot correctly simulate the distributions of different racial, education, and income groups. 

%ChatGPT substantially over-simulates individuals as Blacks or Asians rather than Whites but accurately simulates the proportions of Hispanics. 

GPT estimates also fall short of capturing the proportion of individuals with the lowest level of education while overestimating those with middle or higher levels of education. This finding of the more educated silicon population is corroborated by the findings that GPT tends to show views of a more educated individual~\cite{santurkar2023whose}. The sampling distribution also reveals an underestimation of individuals at the extremes of the income distribution, mirroring patterns observed in social surveys~\cite{NotRandom}. 

More interestingly, when asked directly about the proportion of those racial, educational, or income groups, GPT could provide the correct answer. This highlights the gap between the knowledge acquired and the ability to apply this knowledge to a representative sample, showing little evidence that LLMs can understand.

%Subgroup analysis shows more interesting details. GPT generates more educated respondents especially when they are men or Whites. Therefore, the previous observation of the overestimation of education from the full sample should be more likely driven by Men and Whites, underscoring the bias embedded in the silicon population.

%To test the sensitivity of the generated responses, we employed three role-playing modes: default (no role), the survey respondent, and the survey expert. Despite varying roles, the general conclusions demonstrate consistency and robustness in the sample distributions.

For attitudinal questions, the score distributions of GPT respondents when answering questions about income redistribution follow a bell shape. This is completely different from that of the human responses. This deviation from human responses is a new type of misalignment between language agents and human respondents. The score distribution of GPT respondents also tends to be concentrated in one category with the most human respondents. Examining the mean values of these responses and comparing the GPT agents with the human respondents, We do not find strong support for the more liberal attitudes related to income redistribution and gender roles embedded in ChatGPT as in previous studies~\cite{rutinowski2024self, hartmann2023political}.

 \section{DISCUSSIONS}
Our findings do not provide strong evidence for a significant alignment between responses generated by LLMs (RPLAs) and human responses, contrasting with existing literature on individuals' political attitudes~\cite{argyle2023out,sun2024random}. Instead, our results are more consistent with studies that emphasize the misalignment between LLM outputs and human responses~\cite{Bisbee_Clinton_Dorff_Kenkel_Larson_2024, machinebias2024}. This misalignment arises from the distribution of responses, which tends to follow a bell-shaped curve, showing a high concentration in the most frequently selected category and minimal variation in GPT's responses. This pattern underscores the deterministic nature of machine learning models, which are optimized to align closely with the most accurate predictions of human responses rather than capturing the full spectrum of human attitudes.

The findings align with previous studies noting the deterministic characteristics of LLMs~\cite{song2024good,geng2024largelanguagemodelschameleons}, even though these models have been designed to be less deterministic than other compositional systems~\cite{wu2024newerallmsecurity}. The fundamental design philosophy of LLMs prioritizes providing the most appropriate responses over capturing the significant heterogeneity that exists among different groups, which is a crucial aspect of social surveys.

Future research could investigate other LLMs to determine whether these findings hold true across different models. However, given the inherent differences in design logic between these approaches, we are skeptical that major conclusions would differ when using other LLMs. In fact, a study evaluating the replication abilities of various LLMs found that GPT-3.5 performed the best, providing estimates that were closest to responses to attitudinal questions from the European Social Survey~\cite{geng2024largelanguagemodelschameleons}.

The identified biases in demographic representation and attitudinal distributions suggest that LLMs cannot mirror the complexities of human society. Consequently, studies relying on LLMs as human surrogates should consider these limitations and biases when designing research methodologies and interpreting results. Future research may need to incorporate additional validation steps, such as comparing LLM outputs with diverse human responses or integrating LLM-generated data with traditional survey methods to improve reliability and validity for making inferences about real-world populations.

%We propose that LLMs can be particularly useful for addressing missing values in social surveys, assisting in questionnaire design through their conversational capabilities, and detecting invalid responses from human respondents.

% \subsection{Harmless responses?}
% The silicon paper showed the GPT model reports more harmless responses when topics are sensitive. this kind of shows a strong social desirability phenomenon, which is even stronger in real life. This tendency differs across groups. why is this the case?

% Another issue again is the knowledge of the llm is from the past, thus back looking, undermining the dynamics and potential of change in society, as commented already by xxx and xxx.

% \section{Ethical Considerations}
% \subfile{sections/7_ethics.tex}

\bibliographystyle{ACM-Reference-Format}
\bibliography{sample.bib}

\newpage
 \section{APPENDIX}

\textbf{Appendix: Attitudinal Questions from WVS}

\subsection{WVS subjective questions}
WVS subjective questions and corresponding answer choices.

\begin{minipage}[t]{0.48\textwidth}
\centering
\captionof{table}{WVS Questions about Income Distribution and Gender} 
\begin{tabular}{>{\centering\arraybackslash}m{0.1\linewidth}>{\raggedright\arraybackslash}m{0.1\linewidth}>{\raggedright\arraybackslash}m{0.3\linewidth}>{\raggedright\arraybackslash}m{0.3\linewidth}} \hline Topic& Code& Statements& Answer choices \\ \hline \multirow{2}{*}{\makecell{Income\\inequality}} & Q106& Income should be made more equal. vs. There should be greater incentives for individual efforts.& [1, 10], 1. Income should be made more equal. 10. There should be greater incentives for individual efforts. \\ & Q108& The government should take more responsibility to ensure that everyone is provided for. vs. People should take more responsibility to provide for themselves.& [1, 10], 1. The government should take more responsibility to ensure that everyone is provided for. 10. People should take more responsibility to provide for themselves. \\ \hline \multirow{4}{*}{\makecell{Gender\\inequality}} & Q28& When a mother works for pay, the children suffer& 1. Strongly agree. 2. Agree. 3. Disagree. 4. Strongly disagree. \\ & Q29& On the whole, men make better political leaders than women do.& 1. Strongly agree. 2. Agree. 3. Disagree. 4. Strongly disagree. \\ & Q31& On the whole, men make better business executives than women do.& 1. Strongly agree. 2. Agree. 3. Disagree. 4. Strongly disagree. \\ & Q35& If a woman earns more money than her husband, it's almost certain to cause problems.& 1. Strongly agree. 2. Agree. 3. Neither agree nor disagree. 4. Disagree. 5. Strongly disagree. \\ \hline 
\end{tabular} 
\label{sar2-1} 

\end{minipage}

\end{document}

\textbf{Appendix: Study 1 - The Objective Human Society}

\subsection{Default prompt}
\lstset{
    basicstyle=\ttfamily, 
    breaklines=true, 
    escapeinside={(*@}{@*)}  % 允许在 lstlisting 环境中使用 LaTeX 代码
}
\begin{lstlisting}
prompt\_template = ``````
Your task is to generate a survey dataset representing the US population in 2020. The sample size is 20. The dataset should include 6 demographic variables for a respondent: 
1.sex(male, female)
2.age
3.ethnic group(White; Black or African American; American Indian and Alaska Native; Asian; Hispanic or Latino; Native Hawaiian and Other Pacific Islander; Some Other Race; Two or more races)
4.the highest educational level that you have attained (Less than 9th grade; 9th to 12th grade, no diploma; High school graduate; Some college, no degree; Associate's degree; Bachelor's degree; Graduate or professional degree)
5. Annual household income(Less than $10,000; $10,000 to $14,999; $15,000 to $24,999; $25,000 to $34,999; $35,000 to $49,999; $50,000 to $74,999; $75,000 to $99,999; $100,000 to $149,999; $150,000 to $199,999; $200,000 or more)
6. region(urban; rural)
Here is an example of one data record: [male; 25; White; Some college, no degree; $50,000 to $74,999; urban]. Choose your answers only from the options provided, and please keep your letter case consistent with the example. After generating, only show the data you generated without additional words. The format must be a JSON string representing a multi-dimensional array. Also, make sure that it is an array of arrays with no objects, like in a spreadsheet. Remember, the records should closely reflect the actual population demographic distribution in 2020.
''''''
\end{lstlisting}

\subsection{Survey Expert Role-play prompt}
\lstset{
    basicstyle=\ttfamily, 
    breaklines=true, 
    escapeinside={(*@}{@*)}  % 允许在 lstlisting 环境中使用 LaTeX 代码
}

\begin{lstlisting}
prompt\_template = ``````
You are a statistician and a social survey expert. Your task is to generate a survey dataset representing the US population in 2020. The sample size is 20. The dataset should include 6 demographic variables for a respondent: 
1.sex(male, female)
2.age
3.ethnic group(White; Black or African American; American Indian and Alaska Native; Asian; Hispanic or Latino; Native Hawaiian and Other Pacific Islander; Some Other Race; Two or more races)
4.the highest educational level that you have attained (Less than 9th grade; 9th to 12th grade, no diploma; High school graduate; Some college, no degree; Associate's degree; Bachelor's degree; Graduate or professional degree)
5. Annual household income(Less than $10,000; $10,000 to $14,999; $15,000 to $24,999; $25,000 to $34,999; $35,000 to $49,999; $50,000 to $74,999; $75,000 to $99,999; $100,000 to $149,999; $150,000 to $199,999; $200,000 or more)
6. region(urban; rural)
Here is an example of one data record: [male; 25; White; Some college, no degree; $50,000 to $74,999; urban]. Choose your answers only from the options provided, and please keep your letter case consistent with the example. After generating, only show the data you generated without additional words. The format must be a JSON string representing a multi-dimensional array. Also, make sure that it is an array of arrays with no objects, like in a spreadsheet. Remember, the records should closely reflect the actual population demographic distribution in 2020.
''''''
\end{lstlisting}

\subsection{Survey Respondent Role-play prompt}
\lstset{
    basicstyle=\ttfamily, 
    breaklines=true, 
    escapeinside={(*@}{@*)}  % 允许在 lstlisting 环境中使用 LaTeX 代码
}

\begin{lstlisting}
prompt\_template = ``````
As a survey respondent randomly drawn from the United States in 2020, your task is to answer specific questions in a survey. You will need to answer 6 questions on your:
1.sex(male, female)
2.age
3.ethnic group(White; Black or African American; American Indian and Alaska Native; Asian; Hispanic or Latino; Native Hawaiian and Other Pacific Islander; Some Other Race; Two or more races)
4.the highest educational level that you have attained (Less than 9th grade; 9th to 12th grade, no diploma; High school graduate; Some college, no degree; Associate's degree; Bachelor's degree; Graduate or professional degree)
5. Annual household income(Less than $10,000; $10,000 to $14,999; $15,000 to $24,999; $25,000 to $34,999; $35,000 to $49,999; $50,000 to $74,999; $75,000 to $99,999; $100,000 to $149,999; $150,000 to $199,999; $200,000 or more)
6. region(urban; rural)
Here is an example of one data record: [male; 25; White; Some college, no degree; $50,000 to $74,999; urban]. Choose your answers only from the options provided, and please keep your letter case consistent with the example. After generating, only show the data you generated without additional words. The format must be a JSON string representing a multi-dimensional array. Also, make sure that it is an array of arrays with no objects, like in a spreadsheet. Remember, the records should closely reflect the actual population demographic distribution in 2020.
''''''
\end{lstlisting}

\textbf{Appendix: Study 2 - The Subjective Human Society}

\subsection{Prompt 1}
\lstset{
    basicstyle=\ttfamily, 
    breaklines=true, 
    escapeinside={(*@}{@*)}  % 允许在 lstlisting 环境中使用 LaTeX 代码
}

\begin{lstlisting}
questions = [
    {
    "question_name":"income_first",
    "question_text":"Your characteristics are mentioned in this respondents: {{ respondents }}. Please answer the following question: think of a score of 1 as meaning that incomes should be made more equal, and a score of 10 meaning that there should be greater incentives for individual effort. What score (an integer) between 1 and 10 comes closest to the way you feel?",
    "question_options":[1, 2, 3, 4, 5, 6, 7, 8, 9, 10]
    },
    {
    "question_name":"income_second",
    "question_text":"Your characteristics are mentioned in this respondents: {{ respondents }}. Please answer the following question: think of a score of 1 as meaning that government should take more responsibility to ensure that everyone is provided for, and a score of 10 meaning that people should take more responsibility to provide for themselves. What score (an integer) between 1 and 10 comes closest to the way you feel?",
    "question_options":[1, 2, 3, 4, 5, 6, 7, 8, 9, 10]
    }
]
def create_agent_response(respondent, questions):
    attempts = 3
    responses = {}
    base_prompt = f"""
    Based on the following respondent characteristics:
    - Sex: {respondent['sex']}
    - Age: {respondent['age']}
    - Ethnicity: {respondent['ethnic']}
    - Education: {respondent['education']}
    - Income: {respondent['income']}
    - Chief: {respondent['chief']}
    - Region: {respondent['region']}
    - Party: {respondent['party']}
    """
    for question in questions:
        question_answered = False
        prompt = base_prompt

        for attempt in range(attempts):
            prompt += f"""
                Question: {question['question_text']}
                Options: {', '.join(question['question_options'])}
                Please provide your answer from the options and give a reason.
                Reason:
                Answer:
                """
            data = {
                "model": "gpt-3.5-turbo-0613",
                "messages": [{"role": "user", "content": prompt}],
                "temperature": 1,
                "top_p": 1,
                "logprobs": 10
            }
......

\end{lstlisting}

\subsection{Prompt 2}
\lstset{
    basicstyle=\ttfamily, 
    breaklines=true, 
    escapeinside={(*@}{@*)}  % 允许在 lstlisting 环境中使用 LaTeX 代码
}

\begin{lstlisting}
    
questions = [
    {
    "question_name":"gender_first",
    "question_text":"Your characteristics are mentioned in this respondents: {{ respondents }}. Please answer the following question: How much do you agree with the statement that children suffer when a mother works for pay?",
    "question_options":["Strongly agree","Agree","Disagree","Strongly disagree"]
    },
    {
    "question_name":"gender_second",
    "question_text":"Your characteristics are mentioned in this respondents: {{ respondents }}. Please answer the following question: How much do you agree with the statement that men make better political leaders than women do?",
    "question_options":[ "Strongly agree","Agree","Disagree","Strongly disagree"]
    },
    {
    "question_name":"gender_third",
    "question_text":"Your characteristics are mentioned in this respondents: {{ respondents }}. Please answer the following question: How much do you agree with the statement that men make better business executives than women do?",
    "question_options":[ "Strongly agree","Agree","Disagree","Strongly disagree"]
    },
    {
    "question_name":"gender_fourth",
    "question_text":"Your characteristics are mentioned in this respondents: {{ respondents }}. Please answer the following question: How much do you agree with the statement that if a woman earns more money than her husband, it is almost certain to cause problems?",
    "question_options":[ "Strongly agree","Agree","Neither agree nor disagree","Disagree","Strongly disagree"]
    }
]

def create_agent_response(respondent, questions):
    attempts = 3
    responses = {} 
    base_prompt = f"""
    Based on the following respondent characteristics:
    - Sex: {respondent['sex']}
    - Age: {respondent['age']}
    - Ethnicity: {respondent['ethnic']}
    - Education: {respondent['education']}
    - Income: {respondent['income']}
    - Chief: {respondent['chief']}
    - Region: {respondent['region']}
    - Party: {respondent['party']}
    """
    for question in questions:
        question_answered = False
        prompt = base_prompt

        for attempt in range(attempts):
            prompt += f"""
                Question: {question['question_text']}
                Options: {', '.join(question['question_options'])}
                Please provide your answer from the options and give a reason.
                Reason:
                Answer:
                """
            data = {
                "model": "gpt-3.5-turbo-0613",
                "messages": [{"role": "user", "content": prompt}],
                "temperature": 1,
                "top_p": 1,
                "logprobs": 10
            }
.....

\end{lstlisting}

\subsection{Data processing before sending to API}

\lstset{
    basicstyle=\ttfamily, 
    breaklines=true, 
    escapeinside={(*@}{@*)}  % 允许在 lstlisting 环境中使用 LaTeX 代码
}
\label{appendix2_prompt4}
\begin{lstlisting}

For index, row in df.iterrows():\\
    if index \geq= max\_records:\\
        break  \\
    sex = row['sex']\\
    age = row['age']\\
    ethnic = row['ethnic']\\
    education = row['education']\\
    income = row['income']\\
    chief = row['chief']\\
    region = row['region']\\
    prompt = prompt\_template.format(sex=sex, age=age, ethnic=ethnic, education=education, income=income, chief=chief, region=region, party=party)\\
    data = \{\\
        "model": "gpt-3.5-turbo",\\
        "messages": [{"role": "user", "content": prompt}],\\
        "temperature": 1,\\
        "top\_p": 1,\\
        "logprobs": 10\\
    \}
\end{lstlisting}